\documentclass[12pt]{article}

\usepackage[top=1in, bottom=1in, left=1in, right=1in]{geometry}

\usepackage{type1cm}
\usepackage{makeidx}         
\usepackage{graphicx}        
\usepackage{multicol}        
\usepackage[bottom]{footmisc}

\usepackage{tikz}

\makeindex             

\usepackage{amsmath}
\usepackage{amsfonts}
\usepackage{cite}

\DeclareMathOperator{\grad}{grad}
\DeclareMathOperator{\diver}{div}

\newcommand{\atPoint}[1]{\big|_{#1}}

\newcommand{\totalDiff}[1]{\frac{\mathrm{d}}{\mathrm{d}{#1}}}
\newcommand{\parder}[2]{\frac{\partial #1}{\partial #2}}
\newcommand{\dir}[1]{\partial_{ {#1}}}

\newcommand{\LieDerivative}[2]{\mathcal{L}_{#1}\left(#2\right)}

\newcommand{\LieAlgebra}[1]{\mathfrak{#1}}

\newcommand{\bvec}[1]{\mathbf{#1}}
\newcommand{\entr}{s}
\newcommand{\temp}{T}
\newcommand{\dens}{\rho}
\newcommand{\press}{p}
\newcommand{\energy}{\epsilon}

\newcommand{\thermVars}{\dens,\press,\temp,\entr}
\newcommand{\fullThermVars}{\energy,\thermVars}
\newcommand{\stateSurface}{{L}}
\newcommand{\lagrangianSurface}{\tilde{L}}

\newcommand{\JetSpace}[2]{{J}^{#1}#2}
\newcommand{\systemEk}[1]{\mathcal{E}_{#1}} 
\newcommand{\stress}{\sigma}
\newcommand{\Def}{D}
\newcommand{\visc}{\sigma^{\prime}}
\newcommand{\inner}[2]{{\left\langle#1,#2\right\rangle}_{g}}

\newcommand{\veltg}{\bvec{a_g} }
\newcommand{\velxyz}{\mathrm{V} }

\newtheorem{theorem}{Theorem}
\newtheorem{proposition}{Proposition}

\begin{document}
\title{Differential invariants for flows of fluids and gases}
\author{Anna Duyunova,\\
Institute of Control Sciences of RAS,
\\anna.duyunova@yahoo.com\\
	Valentin Lychagin,\\
	Institute of Control Sciences of RAS,
	valentin.lychagin@uit.no,\\
	Sergey Tychkov,
	\\Institute of Control Sciences of RAS,\\sergey.lab06@ya.ru}
\date{}

\maketitle

\section{Introduction}

The paper is an extended overview of the papers \cite{Duyunova2017-1,Duyunova2017-2,
	Duyunova2017-3,Duyunova2017-4,Duyunova2017-5,Duyunova2017-6,
	Duyunova2017-7}. 
The main extension is a detailed analysis of thermodynamic states,  symmetries, and differential invariants. This analysis is based on consideration of Riemannian  structure  \cite{LychaginWisla18} naturally associated with Lagrangian manifolds that represent thermodynamic states. 
This approach radically changes the description of the thermodynamic part of the symmetry algebra as well as the field  of differential invariants.

The paper is organized as follows.

In Section \ref{sec:thermodynamics} we discuss thermodynamics
in terms of contact and symplectic geometries. The main part of
this approach is a presentation of thermodynamic states as
Lagrangian manifolds equipped with an additional
Riemannian structure. Application of this approach
to fluid motion is new, though the relationship between contact geometry and thermodynamics was well-known since Gibbs \cite{Gibbs1873} and
Carath\'eodory \cite{Caratheodory1909}.
For some modern studies see also \cite{Ruppeiner1995} and
\cite{Bravetti2018}.

In Section \ref{sec:Inviscid} the motion of inviscid media is considered.
We discuss flows of inviscid fluids on different
Riemannian manifolds:
a plane, sphere, and a spherical layer.
Such flows are governed by a generalization of the
Euler equation system. For each of these cases,
we find a Lie algebra of symmetries, provide a classification
of symmetry algebras depending on a thermodynamic state admitted
by media, and describe the field of differential invariants for
the Euler system.

In Section \ref{sec:viscid} the motion of viscid media
on Riemannian manifolds is studied. First, we discuss
a generalization of the Navier--Stokes equations
for an arbitrary oriented Riemannian manifold.
Then for the cases of a plane, space, sphere, and a spherical layer,
we provide classification of symmetry algebras 
with respect to possible thermodynamic states and
give full description for the field of differential invariants.


\section{Thermodynamics}\label{sec:thermodynamics}

Here we consider the media with thermodynamics described by two types of quantities. The first are {\it extensive} quantities: the specific entropy $\entr$, the specific volume $\dens^{-1}$, the specific internal energy $\energy$; and the second are  {\it intensive} quantities: the absolute temperature $\temp>0$ and the pressure $\press$.  

A thermodynamic state of such media is a two-dimensional Legendrian manifold $\stateSurface\subset\mathbb{R}^5 (\fullThermVars)$, a  maximal integral manifold of the differential 1-form 
\[
\theta = d\energy-\temp d\entr-\press\dens^{-2} d\dens,
\]
i.e. a manifold such that the  first law of thermodynamics $\theta\atPoint{\stateSurface}=0$ holds. 






Following \cite{LychaginWisla18} a point $(\fullThermVars)$ on the Legendrian manifold can be considered as a triplet:
the expected value $(\energy, \dens^{-1})$ of a stochastic process of measurement of internal energy and volume, the probabilistic measure corresponding to $(\press, \temp)$ and the information $(-\entr)$, which is given up to a constant.

Since the information $I$
is a positive quantity, the entropy $\entr$ satisfies the inequality $\entr\leq \entr_0$  for a certain
constant $\entr_0$, which, generally speaking, depends on
the nature of a process under consideration.


Let us denote the variance of the stochastic process by $(-\kappa)$.
In terms of the given probabilistic measure and expected values it has form \cite{LychaginWisla18}:
\[
\kappa= d(\temp^{-1})\cdot d\energy - \dens^{-2} d(\press \temp^{-1}) \cdot d\dens.
\]

Thus, by a \textbf{\emph{ thermodynamic state}} we mean a two-dimensional Legendrian submanifold $\stateSurface$ of the contact manifold $(\mathbb{R}^5, \theta)$, such that the quadratic differential form $\kappa$ on the surface $\stateSurface$ is negative definite, i.e.  
\[
\kappa\atPoint{\stateSurface}<0.
\]  

Because the energy can be excluded from the conservation
laws that govern medium motion,
we also eliminate it from our geometrical interpretation of the thermodynamics.

Consider the projection 
\[\varphi : \mathbb{R}^5 \rightarrow \mathbb{R}^4, \quad
\varphi : \left(\fullThermVars\right) \longmapsto
\left(\thermVars \right).
\]
The restriction of the map $\varphi$ on the state surface $\stateSurface$ is a local diffeomorphism
on the image $\lagrangianSurface=\varphi(\stateSurface)$ and
the surface $\lagrangianSurface$ is
an immersed Lagrangian manifold in the symplectic
space $\mathbb{R}^4$ equipped with the structure form 
\[
\Omega = d\entr \wedge d\temp + {\dens^{-2}} d\dens \wedge d\press .
\]

Therefore, the first law of thermodynamics is equivalent to
the condition that
$\lagrangianSurface\subset\mathbb{R}^4$ is a Lagrangian manifold.



The two-dimensional surface $\lagrangianSurface$ in the four-dimensional space can be defined by two equations 
\begin{equation}\label{eq:Therm}
f(\press,\dens,\entr,\temp)=0,\quad
g(\press,\dens,\entr,\temp)=0
\end{equation}
and the condition for the surface $\lagrangianSurface$
to be Lagrangian means vanishing of the Poisson bracket of these functions
\begin{equation}\label{eq:lagr}
[f,g]\atPoint{\lagrangianSurface}=0,
\end{equation}
that in the coordinates $(\press,\dens,\entr,\temp)$ takes the form
\[
[f,g] = \dens^2 \left(f_{\dens}g_{\press} -f_{\press}g_{\dens} \right) + f_{\entr}g_{\temp} - f_{\temp}g_{\entr}.
\]

Thus, {\it the thermodynamic state} can be defined as Lagrangian surface \eqref{eq:Therm} in the four-dimensional symplectic space, such that the condition \eqref{eq:lagr} holds and the symmetric differential form $\kappa$ is negative definite on this surface.  

Note, if the equation of state is given in the form $\energy = \energy(\dens,\entr)$, then the two-dimensional Legendrian manifold $\stateSurface$ can be defined by the structure equations
\begin{equation}\label{ener}
\energy = \energy(\dens,\entr), \quad \temp={\energy}_{\entr}, \quad \press=\dens^2{\energy}_{\dens},
\end{equation}
and the restriction of the form $\kappa$ gives
\[
\kappa\atPoint{\stateSurface} = -\energy_{\entr}^{-1} \left( 
\left( \energy_{\dens\dens} + 2\dens^{-1}\energy_{\dens}\right)  d\dens^2 +2 \energy_{\dens\entr} d\dens\cdot d\entr +\energy_{\entr\entr}
d\entr^2
\right) .
\]

The condition of negative-definiteness this form leads us to the following additional relations 
\[
\left\{
\begin{aligned}
&\energy_{\dens\dens} + 2\dens^{-3}\press >0,\\
&\energy_{\entr\entr} \left( \energy_{\dens\dens} + 2\dens^{-3}\press\right) - \energy_{\dens\entr} ^2>0
\end{aligned}
\right.
\]
 on the function $\energy(\dens,\entr)$ or
\[
\left\{
\begin{aligned}
&\press_{\dens}  >0,\\
&\temp_{\entr}\press_{\dens} - \dens^{2}\temp_{\dens} ^2>0.
\end{aligned}
\right.
\]

\newpage
\section{Compressible inviscid fluids or gases}\label{sec:Inviscid}

In this section we study differential invariants of compressible inviscid fluids or gases. 

The system of differential equations (the Euler system) describing flows on an oriented Riemannian manifold $(M, g)$ consists of the following equations (see \cite{Batchelor2000}  for details):
\begin{equation}\label{eq:E}
\left\{
\begin{aligned}
& \dens(\bvec{u}_t  + \nabla_{\bvec{u}}\bvec{u})=- \grad{\press} +  \bvec{g}\dens  ,\\
& \parder{(\dens\, \Omega_g)}{t} + \LieDerivative{\bvec{u}}{\dens\, \Omega_g} = 0,\\
&\temp\left(  \entr_t + \nabla_{\bvec{u}}s\right)  - \frac{k}{\dens} \Delta_g T =0,
\end{aligned}
\right.
\end{equation}
where the vector field $\bvec{u}$
is the flow velocity, $\press$, $\dens$,
$\entr$, $\temp$ are the pressure, density,
entropy, temperature of the fluid  
respectively, $k$ is the thermal conductivity, which is supposed to be constant, and $\bvec{g}$ is the gravitational acceleration.

Here $\nabla_X$ is the directional covariant Levi--Civita derivative with respect to a vector field $X$, $\mathcal{L}_X$ is the Lie derivative along a vector field $X$,  $\Omega_g$ is the volume form on the manifold $M$, $\Delta_{g}$ is the Laplace--Beltrami operator corresponding to the metric $g$.

The first equation of system \eqref{eq:E} represents the law of momentum conservation in the inviscid medium, the second is the continuity equation,
and the third  is the equation representing the effect of heat conduction in the medium. 

We  consider the following examples of manifold $M$: a plane,
sphere and a spherical layer. 

Note that in all these cases the number of unknown functions is greater than the number of system equations by 2, i.e. the system \eqref{eq:E} is incomplete. We get two additional equations
taking into account the thermodynamics of the medium.

Thus, by the Euler system of differential equations we mean the system~\eqref{eq:E}
extended by two equations of state~\eqref{eq:Therm},
such that the functions $f$ and $g$ satisfy the additional relation~\eqref{eq:lagr} and the form $\kappa$ is negative definite.

Geometrically, we represent this system in the following way.
Consider the bundle of rank $(\dim  M+4)$ 
\[
\pi : \mathbb{R} \times TM \times \mathbb{R}^4  
 \longrightarrow \mathbb{R}\times M,
\]
where 
$
(t, \mathbf{x}, \mathbf{u}, \thermVars) \rightarrow (t,\bar{x})
$
and  $t\in \mathbb{R} $, $\mathbf{x}\in M$, $\mathbf{u}\in T_{\mathbf{x}}M$.
Then the Euler system is a system
of differential equations on sections of the bundle $\pi$.

Note that system~\eqref{eq:Therm} defines the zeroth
order system $\systemEk{0}\subset \JetSpace{0}{\pi}$. 

Denote by $\systemEk{1}\subset \JetSpace{1}{\pi}$ the system of order $\leq 1$ obtained by the first prolongation of the system $\systemEk{0}$
and by the first 2 equations of system~\eqref{eq:E} (Euler's and the continuity equations). 

Let also $\systemEk{2}\subset \JetSpace{2}{\pi}$ be the system of differential
equations of order $\leq 2$ obtained by the first prolongation
of the system $\systemEk{1}$ and the last equation of system~\eqref{eq:E}.

For the case $k\geq 3$, we define $\systemEk{k}\subset\JetSpace{k}{\pi}$
to be the $(k-2)$-th prolongation of the system $\systemEk{2}$.

Note that due to the relations \eqref{eq:Therm} the system $\systemEk{\infty}=\lim\limits_{\longleftarrow}\systemEk{k}$ is a formally
integrable system of differential equations, which we also call the Euler system.


\subsection{2D-flows}

Consider Euler system \eqref{eq:E} on a plane $M=\mathbb{R}^2$ equipped  with the coordinates $(x,y)$ and the standard flat metric $g=dx^2+dy^2$. 

The velocity field of the flow has the form 
$\bvec{u}=u(t,x,y)\,\dir{x} +v(t,x,y)\,\dir{y}$, 
the pressure $\press$, the density $\dens$, the temperature $\temp$ and the entropy $\entr$ are the functions of time and space with the coordinates $(t,x,y)$.

Here we consider the flow without any external force field, so $\bvec{g}={0}$.


\subsubsection{Symmetry Lie algebra}

The symmetry algebra of the Euler system has been found in \cite{Duyunova2017-1}, here we observe the main statements.

First of all, by a symmetry of the PDE system we mean a point symmetry, i.e. a vector field $X$ on the jet space $\JetSpace{0}{\pi}$ such that its second prolongation $X^{(2)}$ is tangent to the submanifold $\systemEk{2}\subset \JetSpace{2}{\pi}$.

To describe the Lie algebra of symmetries of the Euler system, we
consider the Lie algebra $\LieAlgebra{g}$ generated
by the following vector fields on the space $\JetSpace{0}{\pi}$: 
\begin{equation*}
\begin{aligned} 
&X_1 = \dir{ x}, \qquad \phantom{\,y-x\,\dir{ y} + v\,\dir{ u} - u\,\dir{ v} }
X_4 = t\,\dir{ x} + \dir{ u},  \\
&X_2 = \dir{ y}, \qquad \phantom{\,y-x\,\dir{ y} + v\,\dir{ u} - u\,\dir{ v} } 
X_5 = t\,\dir{ y} + \dir{ v} ,  \\
&X_{3} =  y\,\dir{ x} - x\,\dir{ y} + v\,\dir{ u} - u\,\dir{ v}  ,
\qquad X_6 = \dir{t},  \\[2pt]
&X_7 = \dir{ \entr}, \qquad \phantom{T}\qquad
X_{10} =  t\,\dir{ t} + x\,\dir{ x} + y\,\dir{ y} - \entr\,\dir{ \entr} , \\
&X_8 = \dir{ \press} , \qquad  \phantom{T}\qquad
X_{11} = t\,\dir{ t} - u\,\dir{ u} - v\,\dir{ v} - 2\press\,\dir{ \press} + \entr\,\dir{ \entr}, \\
&X_9 =  \temp\dir{ \temp},\qquad \qquad
X_{12} = \press\,\dir{ \press} + \dens\,\dir{ \dens} - \entr\,\dir{ \entr} .
\end{aligned} 
\end{equation*}

Note that this symmetry
algebra consists of pure geometric and thermodynamic parts.

The geometric part $\LieAlgebra{g_{m}}$ is generated by the fields $X_1,\ldots, X_6$. 
Transformations corresponding to the elements of this algebra are generated by the motions, Galilean transformations and the time shift.

In order to describe the pure thermodynamic part of the system symmetry algebra, consider the Lie algebra $\LieAlgebra{h}$ generated by the vector fields
\begin{equation*}
\begin{aligned}
&Y_1 = \dir{ \entr}, \qquad
Y_{3} = \dens\,\dir{ \dens}, \qquad
Y_{5} = \press\,\dir{ \press} ,  \\
&Y_2 = \dir{ \press}, \qquad
Y_4 =  \entr\,\dir{ \entr} , \qquad
Y_6 =  \temp\,\dir{ \temp} .
\end{aligned}
\end{equation*}

Denote by $\vartheta :\LieAlgebra{g}\mapsto\LieAlgebra{h}$ 
the following Lie algebras homomorphism
\begin{equation} \label{theta}
\vartheta(X) = X(\dens)\dir{\dens} + X(\entr)\dir{\entr} + X(\press)\dir{\press} + X(\temp)\dir{\temp},
\end{equation}
where $X\in \LieAlgebra{g}$.

Note that, the kernel of the homomorphism $\vartheta$ is the ideal 
$\LieAlgebra{g_{m}}\subset \LieAlgebra{g}$. 

Let also $\LieAlgebra{h_{t}}$ be the Lie subalgebra of the algebra $\LieAlgebra{h}$
that preserves the thermodynamic state \eqref{eq:Therm}.
\begin{theorem}
	\cite{Duyunova2017-1} A Lie algebra $\LieAlgebra{g_{sym}}$ of symmetries of the Euler system of differential equations on a plane coincides with 
	\[
	\vartheta^{-1}(\LieAlgebra{h_{t}}).
	\]
\end{theorem}

Note that, for the general equation of state, the algebra $\LieAlgebra{h_{t}}=0$, and the symmetry algebra coincides with the Lie algebra $\LieAlgebra{g_{m}}$.

Observe that, usually, the equations of state are neglected, and vector fields like
$f(t)\,\dir{ \press}$ and $g(t)\temp\,\dir{ \temp}$, where $f$ and $g$ are arbitrary functions, are considered as symmetries of the Euler system.

\subsubsection{Symmetry classification of states}\label{sec:classE2}

In this section we classify the thermodynamic states or the Lagrangian surfaces  $\lagrangianSurface$
depending on the dimension of the symmetry algebra $\LieAlgebra{h_{t}}\subset \LieAlgebra{h}$.

We consider one- and two-dimensional symmetry algebras only, because the requirement on the thermodynamic state to have a three or more dimensional symmetry algebra is very strict and leads to the trivial solutions.    

\subsubsection*{States with a one-dimensional symmetry algebra} \label{sec1al}

Let $\dim \LieAlgebra{h_{t}}=1 $ and let $Z = \sum\limits_{i=1}^6 \lambda_i Y_i$ be a basis vector in this algebra.

The state $\lagrangianSurface\subset \mathbb{R}^4 $ is Lagrangian, i.e.
$\Omega\vert_{\lagrangianSurface}=0$, and therefore the vector field $Z$ is tangent
to the surface $\lagrangianSurface$, if and only if the differential 1-form
\[
\iota_{Z}{\Omega} = \frac{\lambda_3}{\dens}\,d\press
- \frac{\lambda_5\press+\lambda_2}{\dens^2}\,d\dens
- \lambda_6 \temp \,d\entr + (\lambda_4\entr+\lambda_1) \,d\temp
\]
vanishes on the surface $\lagrangianSurface$.

In other words, the surface $\lagrangianSurface$ is the solution of the following system
of differential equations
\[
\left\{
\begin{aligned}
&\Omega\vert_{\lagrangianSurface} =0, \\
&(\iota_{Z}{\Omega})\vert_{\lagrangianSurface} = 0.
\end{aligned} \right.
\]
In terms of specific energy \eqref{ener} the last system has the following form
\begin{equation*}
\left\{
\begin{aligned}
&\lambda_3\dens\,{\energy_{\dens\dens}} + ( \lambda_4\entr+\lambda_1 ){\energy_{\dens\entr}}  + \left(  2\lambda_3 - \lambda_5 \right){\energy_{\dens}}  - \frac{\lambda_2}{\dens^2} =0, \\
&(\lambda_4\entr+\lambda_1){\energy_{\entr\entr}} + \lambda_3\dens\,{\energy_{\dens\entr}} -\lambda_6\,{\energy}_{ \entr} =0.
\end{aligned} \right. 
\end{equation*}

It is easy to check that the bracket of these last two equations (see \cite{Kruglikov2002}) vanishes, and therefore the system is formally integrable and compatible.

In order to solve the last system we reduce its order and get the equivalent system
\begin{equation*}
\left\{
\begin{aligned}
& \lambda_3\dens\,{\energy}_{\dens} + (\lambda_4\entr+\lambda_1){\energy}_{ \entr} + (\lambda_3-\lambda_5)\energy + \dfrac{\lambda_2}{\dens} + f(\entr) = 0, \\
& \lambda_3\dens\,{\energy}_{\dens} + (\lambda_4\entr+\lambda_1){\energy}_{ \entr} - (\lambda_6+\lambda_4)\energy + g(\dens)=0,
\end{aligned} \right. 
\end{equation*}
where $f(\entr)$ and $g(\dens)$ are some differentiable functions. 

Below we list solutions of the system under the assumption of parameters $\lambda$ generality. The more detailed description can be found in \cite{Duyunova2017-1}.

In the general case, when {$ \lambda_6+\lambda_4-\lambda_5+\lambda_3 \neq 0$}, solving the last system we find 
\begin{equation*}
\press =C_1\dens^{\frac{\lambda_5}{\lambda_3}}     - \frac{\lambda_2 }{\lambda_5}  ,\quad
\temp =C_2(\lambda_4\entr+\lambda_1)^{\frac{\lambda_6}{\lambda_4}}   ,
\end{equation*}
where $C_1, C_2$ are constants.

Moreover, the negative definiteness of the quadratic differential form $\kappa$ on the surface $\lagrangianSurface$ leads to the relations
\[
\frac{\lambda_4\entr+\lambda_1}{\lambda_6} >0, \quad
\frac{C_1\lambda_5}{\lambda_3} >0
\]
for all $\entr \in(-\infty,\entr_0]$.
\begin{theorem}
	The thermodynamic states admitting a one-dimensional symmetry algebra  have the form 
	\begin{equation*}
	\press =C_1\dens^{\frac{\lambda_5}{\lambda_3}}     - \frac{\lambda_2 }{\lambda_5}  ,\quad
	\temp =C_2(\lambda_4\entr+\lambda_1)^{\frac{\lambda_6}{\lambda_4}}   ,
	\end{equation*}
	where the constants defining the symmetry algebra satisfy inequalities 
	\[
	\entr_0<-\frac{\lambda_1}{\lambda_4}, \quad
	C_1>0, \quad  \frac{\lambda_5 }{\lambda_3}>0, \quad  \frac{\lambda_2 }{\lambda_5}<0,
	\]
	and besides they must meet one of the following conditions:
	\begin{enumerate}
		\item if $\frac{\lambda_6}{\lambda_4}$ is irrational, then  $\lambda_4<0$, $\lambda_6>0$, $C_2>0$;
		\item if $\frac{\lambda_6}{\lambda_4}$ is rational, then $\frac{\lambda_6}{\lambda_4}<0$ (i.e. $\frac{\lambda_6}{\lambda_4}=-\frac{m}{k}$ ) and
		\begin{enumerate}
			\item if $k$ is even, then $\lambda_4<0$, $C_2>0$;
			\item if $k$ is odd and $m$ is even, then $C_2>0$;
			\item if $k$ is odd and $m$ is odd, then $C_2\lambda_4<0$.
		\end{enumerate}
	\end{enumerate}
\end{theorem}

\subsubsection*{States with a two-dimensional non-commutative symmetry algebra}

Let $\LieAlgebra{h_t} \subset \LieAlgebra{h}$ be a non-commutative two-dimensional Lie subalgebra.
Then $[\LieAlgebra{h}, \LieAlgebra{h}]\supset[\LieAlgebra{h_t},\LieAlgebra{h_t}] =\langle Y_1,Y_2\rangle $.
Therefore, any non-zero vector $A=\alpha_0 Y_1+\beta_0 Y_2\in\LieAlgebra{h_t}$ can be chosen as one of the basis vectors.
The second basis vector $B$ in the subalgebra may be chosen such that $\left[ A,B\right] = A$.
Let $B=\sum\limits_{i=1}^6 \gamma_iY_i$, then the condition $\left[ A,B\right] = A$ gives two relations
\begin{equation*} 
\alpha_0(\gamma_4-1)=0, \quad \beta_0(\gamma_5-1)=0.
\end{equation*}

Restriction of the forms $\iota_{A}{\Omega}$ and $\iota_{B}{\Omega}$ 
on the surface $\stateSurface$ leads us to the following system of differential equations:
\begin{equation*}
\left\{
\begin{aligned}
& \alpha_0{\energy}_{\entr\entr} =0 , \\
& \alpha_0\dens^2{\energy}_{\dens\entr} -{\beta_0} =0 , \\
& \gamma_3\dens{\energy}_{\dens \entr} + \left( \gamma_4s+\gamma_1\right) {\energy}_{\entr\entr} -\gamma_6{\energy}_{\entr} =0,\\
&\gamma_3\left( \dens{\energy}_{ \dens\dens}+2{\energy}_{ \dens}\right) + \left( \gamma_4s+\gamma_1\right) {\energy}_{\dens\entr} - \gamma_5{\energy}_{\partial \dens} -\frac{\gamma_2}{\dens^2}  =0.
\end{aligned} \right. 
\end{equation*}

Note that from the first two equations of this system follows that $\alpha_0 \neq0$.

Computing brackets \cite{Kruglikov2002} we get that this system is integrable if
\begin{equation*}
\beta_0(\gamma_4-\gamma_5)=0, \quad \beta_0(\gamma_3\gamma_5+\gamma_4\gamma_6)=0.
\end{equation*}

Then solving this system for the case  $\beta_0=0$ and  $\gamma_4=1$
we have $\temp =  0$ which is not sensible from the physical point of view.

For the case $\gamma_4=1$ and $\gamma_5=1$ we get
\begin{equation*}
\press = C\dens^{\frac{1}{\gamma_3}} + \frac{\beta_0}{\alpha_0}\left(\entr+ \gamma_1\right)  - {\gamma_2} , \quad
\temp =  -\frac{\beta_0}{\alpha_0\dens}  ,
\end{equation*}

but the condition on the form $\kappa$  gives 
\[
\frac{C}{\gamma_3} >0, \quad  -\frac{1}{\dens^2}>0.
\]

So there are no thermodynamic states that admit a two-dimensional non-commutative symmetry algebra.

\subsubsection*{States with a two-dimensional commutative symmetry algebra}

Let now $\LieAlgebra{h_t} \subset \LieAlgebra{h}$ be a commutative two-dimensional Lie subalgebra,
and let $A=\sum\limits_{i=1}^6 \alpha_iY_i$, $B=\sum\limits_{i=1}^6 \beta_iY_i$
be basis vectors in the algebra $\LieAlgebra{h_t}$.

Then the condition $[A,B]=0$ gives the following relations on $\alpha$'s and $\beta$'s:
\begin{equation} \label{commut1}
\alpha_1\beta_4 - \alpha_4\beta_1 = 0, \quad \alpha_2\beta_5 - \alpha_5\beta_2 = 0 .
\end{equation}

Then, as above, restriction of the forms $ \iota_{A}{\Omega}$ and $\iota_{B}{\Omega}$
on the state surface $\lagrangianSurface$ leads us to the following system of differential equations:
\begin{equation*}
\left\{
\begin{aligned}
&\alpha_3\dens\,{\energy}_{\dens\dens} + ( \alpha_4s+\alpha_1 ){\energy}_{\dens \entr}  + \left(  2\alpha_3 - \alpha_5 \right){\energy}_{\dens}  - \frac{\alpha_2}{\dens^2} =0, \\
&\beta_3\dens\,{\energy}_{\dens\dens} + ( \beta_4s+\beta_1 ){\energy}_{\dens \entr}  + \left(  2\beta_3 - \beta_5 \right){\energy}_{\dens}  - \frac{\beta_2}{\dens^2} =0, \\
&(\alpha_4s+\alpha_1){\energy}_{\entr\entr} + \alpha_3\dens\,{\energy}_{\dens\entr} -\alpha_6\,{\energy}_{\entr} =0,\\
&(\beta_4s+\beta_1){\energy}_{\entr\entr} + \beta_3\dens\,{\energy}_{\dens \entr} -\beta_6\,{\energy}_{\entr} =0.
\end{aligned} \right. 
\end{equation*}

The formal integrability condition for this system has the form
\begin{equation*}
{(\beta_5-5\beta_3)(\alpha_2\beta_5-\alpha_5\beta_2)}=0,
\end{equation*}
which is satisfied due to relations \eqref{commut1}.

Therefore, this system is integrable, and for all $\alpha$'s and $\beta$'s. In most of cases this system has the ``nonphysical'' solution of the form $\energy = C_1\dens^{-1} +C_2$.
For the special case, for example,
\begin{equation}\label{case1}
\alpha_1= \frac{\alpha_4\beta_1}{\beta_4},\quad
\alpha_2= \frac{\alpha_5\beta_2}{\beta_5} \quad \text{and} \quad
\left\{
\begin{aligned}
& \alpha_3 = \alpha_5-\alpha_4-\alpha_6 , \\
& \beta_3 = \beta_5-\beta_4-\beta_6
\end{aligned} \right. 
\end{equation}
we have the following expressions for the pressure and the temperature 
\begin{equation*}
\press =  C_1\dens^{\frac{\varsigma_2}{\varsigma_1+\varsigma_2}}
(\beta_4\entr+\beta_1)^{\frac{\varsigma_3 +\varsigma_2}{\varsigma_1+\varsigma_2}}
- \frac{\beta_2}{\beta_5} ,
\quad
\temp =  C_2\dens^{\frac{\varsigma_2}{\varsigma_1+\varsigma_2}}
(\beta_4\entr+\beta_1)^{\frac{\varsigma_3 +\varsigma_2}{\varsigma_1+\varsigma_2}} 
\dens^{-1} 	(\beta_4\entr+\beta_1)^ {-1},
\end{equation*}
where 
\begin{equation*}
\varsigma_1 = \alpha_6\beta_4-\alpha_4\beta_6, \quad
\varsigma_2 = \alpha_4\beta_5-\alpha_5\beta_4, \quad
\varsigma_3 = \alpha_6\beta_5 -\alpha_5\beta_6.
\end{equation*}

And negative definiteness of the form $\kappa$ leads to the relations
\[
\frac{\varsigma_2(\beta_4\entr+\beta_1)}{\varsigma_3\beta_4} >0,
\quad 
\frac{-\varsigma_1}{(\varsigma_1+\varsigma_2)(\varsigma_2+\varsigma_3)}>0.
\]
\begin{theorem} In the general case, there are no physically applicable
thermodynamic states, which admit a two-dimensional commutative symmetry algebra.
	    
	For the special case \eqref{case1}, the thermodynamic states admitting a two-dimensional commutative symmetry algebra have the form 
	\begin{equation*}
	\press =  C_1\dens^{\frac{\varsigma_2}{\varsigma_1+\varsigma_2}}
	(\beta_4\entr+\beta_1)^{\frac{\varsigma_3 +\varsigma_2}{\varsigma_1+\varsigma_2}}
	- \frac{\beta_2}{\beta_5} ,
	\quad
	\temp =  C_2\dens^{\frac{-\varsigma_1}{\varsigma_1+\varsigma_2}}
	(\beta_4\entr+\beta_1)^{\frac{\varsigma_3 -\varsigma_1}{\varsigma_1+\varsigma_2}},
	\end{equation*}
	where the constants defining the symmetry algebra satisfy inequalities
	\[
	\entr_0<-\frac{\beta_1}{\beta_4}, \quad
	\frac{\beta_2}{\beta_5}<0, \quad
	 \frac{\varsigma_2 }{\varsigma_3}<0,
	 \quad 
	 \frac{\varsigma_1}{(\varsigma_1+\varsigma_2)(\varsigma_2+\varsigma_3)}<0,
	\]
	and besides they must meet one of the conditions:
	\begin{enumerate}
		\item if $\frac{\varsigma_3 +\varsigma_2}{\varsigma_1+\varsigma_2}$ is irrational, then  $\beta_4<0$, $C_1>0$, $C_2>0$;
		\item if $\frac{\varsigma_3 +\varsigma_2}{\varsigma_1+\varsigma_2}$ is rational (i.e. $\frac{\varsigma_3 +\varsigma_2}{\varsigma_1+\varsigma_2}=\pm\frac{m}{k}$ ), then 
		\begin{enumerate}
			\item if $k$ is even, then $\beta_4<0$, $C_1>0$, $C_2>0$;
			\item if $k$ is odd and $m$ is even, then $C_1\beta_4<0$,  $C_2>0$;
			\item if $k$ is odd and $m$ is odd, then $C_2\beta_4<0$, $C_1>0$.
		\end{enumerate}
	\end{enumerate}
\end{theorem}

\subsubsection{Differential invariants}\label{sec:invE2d}

We consider two group actions on the Euler equation $\systemEk{}$. The first one is the prolonged  action of the group generated
by the action of the Lie algebra $\LieAlgebra{g_{m}}$. The second action is the action generated by the prolongation
of the action of the Lie algebra $\LieAlgebra{g_{sym}}$. 

First of all, observe that fibers of the projection $\systemEk{k}\rightarrow \systemEk{0}$ are irreducible algebraic manifolds.

Then we say that 
a function $J$ on the manifold $\systemEk{k}$ is a \textit{kinematic differential invariant of order} $\leq k$ if 
\begin{enumerate}
	\item $J$ is a rational function along fibers of the projection $\pi_{k,0}:\systemEk{k}\rightarrow \systemEk{0}$,
	\item $J$ is invariant with respect to the prolonged action of the Lie algebra $\LieAlgebra{g_{m}}$, i.e. 
	\begin{equation} \label{dfinv}
	X^{(k)}(J)=0,
	\end{equation}
	for all $X\in \LieAlgebra{g_{m}}$.
\end{enumerate}
Here we denote by $X^{(k)}$ the $k$-th prolongation of a vector field $X\in \LieAlgebra{g_{m}}$.

We say also that the kinematic invariant is \textit{an Euler invariant} if condition \eqref{dfinv} holds for all $X \in \LieAlgebra{g_{sym}}$.

We say that a point $x_k\in \systemEk{k}$ and the corresponding orbit $\mathcal{O}(x_k)$ ($\LieAlgebra{g_{m}}$ or $\LieAlgebra{g_{sym}}$-orbit) are \textit{regular}, if there are exactly $m=\mathrm{codim}\, \mathcal{O}(x_k) $ independent  invariants (kinematic or Euler) in a neighborhood of this orbit. 

Thus, the corresponding point on the quotient $\systemEk{k} /{\LieAlgebra{g_{m}}}$ or $\systemEk{k} /{\LieAlgebra{g_{sym}}}$ 
is smooth, and these independent invariants  (kinematic or Euler) can serve as
local coordinates in a neighborhood of this point.

Otherwise, we say that the point and the corresponding orbit are \textit{singular}. 

It is worth to note that the Euler system together with the symmetry algebras
$ \LieAlgebra{g_{m}}$ or $\LieAlgebra{g_{sym}}$ satisfies the conditions
of Lie-Tresse theorem (see \cite{KL}), and therefore
the kinematic 
and Euler differential invariants separate
regular $ \LieAlgebra{g_{m}}$ and $\LieAlgebra{g_{sym}}$
orbits on the Euler system $\systemEk{}$ correspondingly. 

By a $\LieAlgebra{g_{m}}$ or $\LieAlgebra{g_{sym}}$-invariant derivation we mean a total derivation
\[
\nabla=A\totalDiff{t}+B\totalDiff{x}+C\totalDiff{y}
\]
that commutes with prolonged action of algebra $\LieAlgebra{g_{m}}$  or $\LieAlgebra{g_{sym}}$.
Here $A$, $B$, $C$ are rational functions on the prolonged equation $\systemEk{k}$ for some $k\geq 0$.

\subsubsection*{The field of kinematic invariants}

First of all, observe that the functions
\[
\dens, \quad \entr
\]
(as well as $\press$ and $\temp$) on the equation $\systemEk{0}$ are $\LieAlgebra{g_{m}}$-invariants. 

Straightforward computations using DifferentialGeometry package by I. Anderson \cite{Anderson2016}
in Maple show that the following functions are the first order kinematic invariants:
\begin{equation*}
\begin{aligned}
&J_1=u_x+v_y, \qquad \,
J_5= \dens_x\entr_y-\dens_y\entr_x ,\\
&J_2=u_y-v_x  ,\qquad  \,
J_6= \entr_t + \entr_x u + \entr_y v,  \\
&J_3 = \dens_x^2 + \dens_y^2, \qquad 
J_7=  \dens_x(\dens_xu_x+\dens_yu_y) + \dens_y(\dens_xv_x+\dens_yv_y), \\
&J_4 = \entr_x^2 + \entr_y^2, \, \qquad  \,
J_{8}= \entr_x(\dens_xu_x+\dens_yu_y) + \entr_y(\dens_xv_x+\dens_yv_y) .
\end{aligned}
\end{equation*}

It is easy to check that the codimension of the regular $\LieAlgebra{g_{m}}$-orbits
on $\systemEk{1}$ is equal to 10.
\begin{proposition}
	The singular points belong to the union of two sets: 
	\begin{equation*}
	\Upsilon_1 = \lbrace\, u_x - v_y = 0, \,\, u_y + v_x = 0, \,\,
	u_t=v_t=\dens_x=\dens_y=\entr_x=\entr_y  = 0 \,\rbrace,
	\end{equation*}
	\[
	\Upsilon_2 = \lbrace\,J_3J_5(J_3J_4-J_5^2)=0 \,\rbrace.
	\]
	The set $\Upsilon_1$ contains singular points that have five-dimensional
	singular orbits.
	The set $\Upsilon_2$ contains points where differential invariants $J_1,J_2,\ldots,J_{8}$ are dependent.
\end{proposition}

The proofs of the following theorems can be found in \cite{Duyunova2017-1}.
\begin{theorem} \cite{Duyunova2017-1}
	The field of the first order kinematic invariants 
	is generated by the invariants $\dens,\entr,J_1,J_2,J_3,\ldots,J_{8}$. 
	These invariants separate the regular $\LieAlgebra{g_m}$-orbits.
\end{theorem}
\begin{theorem} \cite{Duyunova2017-1}
	The derivations 
	\[
	\nabla_1 = \totalDiff{t}+u\totalDiff{x}+v\totalDiff{y},\quad
	\nabla_2 = \dens_x\totalDiff{x}+\dens_y\totalDiff{y},\quad
	\nabla_3 = \entr_x\totalDiff{x}+\entr_y\totalDiff{y}
	\]
	are $\LieAlgebra{g_{m}}$-invariant. They are linearly independent if 
	\[
	\dens_x\entr_y-\dens_y\entr_x\neq 0.
	\]
\end{theorem}

The bundle $\pi_{2,1}:\systemEk{2}\rightarrow \systemEk{1}$ has rank 14,
and by applying the derivations $\nabla_1, \nabla_2, \nabla_3$ to the kinematic invariants $J_1,J_2,\ldots,J_{8}$ we get 24 kinematic invariants. Straightforward computations show that among these invariants 14 are always independent (see  \textit{http://d-omega.org}).

Moreover, starting with the order $k=1$ dimensions of regular orbits are equal to $\dim \LieAlgebra{g_m}=6$ and all equations $\systemEk{k}$, $k\geq3$, are the prolongations of $\systemEk{2}$. 

Therefore, if we denote by $H(k)$ the Hilbert function of the $\LieAlgebra{g_m}$-invariants field,
i.e. $H(k)$ is the number of independent
invariants of pure order $k$ (see \cite{KL} for details), then $H(k) = 5k +4$ for $k\geq 2$, and $H(0)=2$,  $H(1)=8$.

The corresponding Poincar\'{e} function is equal to
\begin{equation*}
P(z) = \frac{2+4z-z^3}{(1-z)^2}.
\end{equation*}

Summarizing, we get the following result.
\begin{theorem} \cite{Duyunova2017-1}
	The field of the kinematic invariants is generated by the invariants $\dens,\entr$ of order zero,
	by the invariants $J_1,J_2,\ldots,J_{8}$ of order one and by the invariant derivations  $\nabla_1, \nabla_2, \nabla_3$.
	This field separates the regular orbits.
\end{theorem}

\subsubsection*{The field of Euler invariants }

Let us consider the case when the thermodynamic state admit a one-dimensional symmetry algebra generated by the vector field
\begin{equation*}
A =  \xi_1 X_7 + \xi_2 X_8 + \xi_3 X_9 + \xi_4 X_{10} + \xi_5 X_{11} + \xi_6 X_{12} .
\end{equation*}
Note that, the $\LieAlgebra{g_{m}}$-invariant derivations $\nabla_1$, $\nabla_2$, $\nabla_3$ do not commute with the thermodynamic symmetry $A$. 

Moreover, the action of the thermodynamic vector field $A$ on the field of kinematic invariants is given by the following derivation
\[
\xi_6\dens\dir{\dens} + \left(\xi_1 - \entr(\xi_4-\xi_5+\xi_6) \right) \dir{\entr} - J_1(\xi_4+\xi_5)\dir{J_1} - J_2(\xi_4+\xi_5)\dir{J_2} -
\]
\[
2J_3(\xi_4-\xi_6)\dir{J_3} - 2J_4(2\xi_4-\xi_5+\xi_6)\dir{J_4} - J_5(3\xi_4-\xi_5)\dir{J_5} - J_6(2\xi_4+\xi_6)\dir{J_6} -
\]
\[
 J_{7}(3\xi_4+\xi_5-2\xi_6)\dir{J_{7}}
 -4\xi_4J_8\dir{J_8}.
\]

Therefore, finding the first integrals of this vector field we get the basic Euler invariants of the first order. 
\begin{theorem} \cite{Duyunova2017-1}
	The field of the Euler differential invariants for thermodynamic states
	admitting a one-dimensional symmetry algebra is generated by
	the differential invariants
	\[
	\frac{J_1}{J_6}\left( \entr - \frac{\xi_1}{\xi_4-\xi_5+\xi_6}  \right) , \quad 
	J_1 \dens^{\frac{\xi_4+\xi_5}{\xi_6}}, \quad
	\frac{J_2}{J_1}, \quad 
	\frac{J_3}{\dens^3 J_6},
	\]
	\[
	\frac{J_4 J_1^2}{\dens J_6^3}, \quad
	\frac{J_5J_1}{J_8}, \quad 
	J_6 \dens^{\frac{2\xi_4}{\xi_6}+1}, \quad 
	\frac{J_{7}}{J_1J_3}, \quad
	\frac{J_8}{\dens^2J_6^2}
	\]
	of the first order and by the invariant derivations 
	\[
	\dens^{\frac{\xi_4+\xi_5}{\xi_6}} \nabla_1,\quad
	\dens^{\frac{2\xi_4}{\xi_6}-1} \nabla_2,\quad
	\dens^{\frac{3\xi_4-\xi_5}{\xi_6} +1} \nabla_3 .
	\] 
	This field separates the regular orbits.
\end{theorem}

Now consider the case when the thermodynamic state
admits a commutative two-dimen\-sio\-nal symmetry
algebra generated by the vector fields 
$A=\sum\limits_{i=1}^6 \mu_iX_{i+6}$
and $B=\sum\limits_{i=1}^6 \eta_iX_{i+6}$
such that $\mu$'s and $\eta$'s satisfy relations 

\begin{equation*}
\left\{
\begin{aligned}
& \eta_1\mu_4-\eta_4\mu_1-\eta_1\mu_5+\eta_5\mu_1+\eta_1\mu_6-\eta_6\mu_1=0, \\
& 2\eta_2\mu_5-2\eta_5\mu_2-\eta_2\mu_6+\eta_6\mu_2=0.
\end{aligned} \right. 
\end{equation*}

Using similar computations we get the following result.
\begin{theorem} \cite{Duyunova2017-1}
	The field of the Euler differential invariants for thermodynamic states admitting
	a com\-mu\-ta\-tive two-dimensional symmetry algebra is generated by differential invariants 
	\[
     J_1 \dens^{\frac{ \varsigma_1+\varsigma_2-2\varsigma_3  }{\varsigma_2-\varsigma_1} }  ((\mu_4-\mu_5+\mu_6)\entr -\mu_1)^{\frac{\varsigma_2+\varsigma_1}{\varsigma_2-\varsigma_1} }  , \quad
	\frac{J_2}{J_1}, \quad
	\frac{J_3}{\dens^3J_1((\mu_4-\mu_5+\mu_6)\entr -\mu_1)} , 
	\]
	\[
	\frac{\dens^8J_1^2J_4}{J_3^3}, \quad
	{\frac{\dens^4J_1J_5}{J_3^2}}, \quad
	\frac{\dens^3J_6}{ J_3}, \quad
	\frac{\dens^4J_7}{J_3^2}, \quad
	\frac{J_{8}}{J_1J_3}
	\]
	of the first order and by the invariant derivations 
	\[
    \dens^{\frac{ \varsigma_1+\varsigma_2-2\varsigma_3  }{\varsigma_2-\varsigma_1} }  ((\mu_4-\mu_5+\mu_6)\entr -\mu_1)^{\frac{\varsigma_2+\varsigma_1}{\varsigma_2-\varsigma_1} }
     \, \nabla_1,
   \quad
	\dens^{\frac{ 3\varsigma_1-\varsigma_2-2\varsigma_3  }{\varsigma_2-\varsigma_1} } 
	((\mu_4-\mu_5+\mu_6)\entr-\mu_1)^{\frac{2\varsigma_1}{\varsigma_2-\varsigma_1} }
	 \, \nabla_2,
	 \]
	 \[
	\dens^{\frac{ 2\varsigma_1-2\varsigma_3  }{\varsigma_2-\varsigma_1}}  ((\mu_4-\mu_5+\mu_6)\entr -\mu_1)^{\frac{3\varsigma_1-\varsigma_2}{\varsigma_2-\varsigma_1} } 	
	\, \nabla_3,
	\]
	where
	\begin{equation*}
	\varsigma_1 = \eta_4\mu_6-\eta_6\mu_4, \quad
	\varsigma_2 = \eta_5\mu_6-\eta_6\mu_5, \quad
	\varsigma_3 = \eta_4\mu_5-\eta_5\mu_4.
	\end{equation*}
	This field separates the regular orbits.
\end{theorem}

Note that these theorems are valid for general $\xi$'s. The special cases are considered in \cite{Duyunova2017-1}.

\subsection{Flows on a sphere}
In this section we consider Euler system \eqref{eq:E} on a two-dimensional unit sphere $M=S^2$  with the metric $g=\sin^2y\,dx^2+dy^2$ in the spherical coordinates.

The velocity field of the flow has the form 
$\bvec{u}=u(t,x,y)\,\dir{x} +v(t,x,y)\,\dir{y}$, 
the pressure $\press$, the density $\dens$, the temperature $\temp$ and the entropy $\entr$ are the functions of time and space with the coordinates $(t,x,y)$.

Here we consider the flow without any external force field, so $\bvec{g}={0}$.

\subsubsection{Symmetry Lie algebra}

As in the previous section, to describe the Lie algebra of symmetries, we consider the Lie algebra $\LieAlgebra{g}$ generated
by the following vector fields on the manifold $\JetSpace{0}{\pi}$: 
\begin{equation*} 
\begin{aligned} 
&X_1 = \dir{ t}, \qquad
X_2 = \dir{ x}, \qquad  \\
&X_{3} = \frac{\cos x}{\tan y} \,\dir{ x} + \sin x \,\dir{ y} -
\left( \frac{\sin x}{\tan y}\,u + \frac{\cos x}{\sin^2 y} \,v \right) \dir{ u} + u\cos x \,\dir{ v}  ,  \\
&X_{4} =   \frac{\sin x}{\tan y} \,\dir{ x} - \cos x \,\dir{ y} +
\left( \frac{\cos x}{\tan y}\,u - \frac{\sin x}{\sin^2 y} \,v \right) \dir{ u} + u\sin x \,\dir{ v}  ,  \\
&X_{5} =  \dir{ \entr} , \qquad  X_{6} =  \dir{ \press} , \qquad
X_{7} = \temp\,\dir{ \temp}  ,\\
&X_{8} = t\,\dir{ t} - u\,\dir{ u} - v\,\dir{ v} + 2\dens\,\dir{ \dens} - \entr\,\dir{ \entr}, \\
&X_{9} = \press\,\dir{ \press} + \dens\,\dir{ \dens} - \entr\,\dir{ \entr} .
\end{aligned} 
\end{equation*}

Consider the pure geometric and thermodynamic parts of this symmetry algebra.

The geometric part $\LieAlgebra{g_{m}}=\langle X_1,X_2,X_3,X_4\rangle$  represents by the symmetries with respect to a group of sphere motions and
time shifts, i.e. $\LieAlgebra{g_{m}} = \LieAlgebra{so}(3,\mathbb{R})\oplus \mathbb{R}$, and  $\LieAlgebra{g_{m}}=\ker\vartheta$. 

To describe the thermodynamic part of the symmetry algebra, we
denote by $\LieAlgebra{h}$ the Lie algebra generated by the vector fields
\begin{equation*}
\begin{aligned}
&Y_1 = \dir{ \entr}, \qquad
Y_2 = \dir{ \press}, \qquad
Y_3 =  \temp\,\dir{ \temp} ,\\
&Y_4 =  2\dens\,\dir{ \dens} - \entr\,\dir{ \entr}, \qquad
Y_{5} = \press\,\dir{ \press} - \dens\,\dir{ \dens}  .
\end{aligned}
\end{equation*}

This is a solvable Lie algebra with the following structure
\[
\left[ Y_1,Y_4\right] =-Y_1, \qquad  \left[ Y_2,Y_5\right] =Y_2.
\]

As above, let $\LieAlgebra{h_{t}}$ be the Lie subalgebra of algebra $\LieAlgebra{h}$
that preserves thermodynamic state \eqref{eq:Therm}.
\begin{theorem} \cite{Duyunova2017-2}
	The Lie algebra $\LieAlgebra{g_{sym}}$ of point symmetries of the Euler system of differential equations on a two-dimensional unit sphere coincides with 
	\[
	\vartheta^{-1}(\LieAlgebra{h_{t}}).
	\]
\end{theorem}

\subsubsection{Symmetry classification of states}

In this section we classify the thermodynamic states or Lagrangian surfaces  $\lagrangianSurface$ (compare with the previous section)
depending on the dimension of the symmetry algebra $\LieAlgebra{h_{t}}\subset \LieAlgebra{h}$.

We consider one- and two-dimensional symmetry algebras only.

\subsubsection*{States with a one-dimensional symmetry algebra} 

Let $\dim \LieAlgebra{h_{t}}=1 $ and let $Z = \sum\limits_{i=1}^5 \lambda_i Y_i$ be a basis vector in this algebra,
then the differential 1-form $\iota_{Z}{\Omega}$ has the form
\[
\iota_{Z}{\Omega} = \frac{2\lambda_4-\lambda_5}{\dens}\,d\press
- \frac{\lambda_5\press+\lambda_2}{\dens^2}\,d\dens
- \lambda_3 \temp \,d\entr + (\lambda_1-\lambda_4\entr) \,d\temp,
\]
and the surface $\lagrangianSurface$ can be found from the following PDE system  
\begin{equation}\label{qwe1}
\left\{
\begin{aligned}
&(2\lambda_4-\lambda_5)\dens\,{\energy}_{\dens\dens} + ( \lambda_1-\lambda_4\entr ){\energy}_{\dens \entr}  + \left(  4\lambda_4 - 3\lambda_5 \right){\energy}_{\dens}  - \frac{\lambda_2}{\dens^2} =0, \\
&(\lambda_1-\lambda_4\entr){\energy}_{\entr \entr} + (2\lambda_4-\lambda_5)\dens\,{\energy}_{\dens \entr} -\lambda_3\,{\energy}_{ \entr} =0.
\end{aligned} \right. 
\end{equation}

It is easy to check that the bracket of these two equations (see \cite{Kruglikov2002}) vanishes, and therefore the system is formally integrable and compatible.

Below we list solutions of this system under the assumption of parameters $\lambda$ generality. A more detailed description may be found in \cite{Duyunova2017-1}, \cite{Duyunova2017-2}.

Solving the last system in case  $\lambda_3+\lambda_4-2\lambda_5 \neq 0$,  we find the following expressions for the pressure and the temperature: 
\begin{equation*}
\press =C_1\dens^{\frac{\lambda_5}{2\lambda_4-\lambda_5}}     - \frac{\lambda_2 }{\lambda_5}  ,\quad
\temp =C_2(\lambda_1-\lambda_4\entr)^{-\frac{\lambda_3}{\lambda_4}}   ,
\end{equation*}
where $C_1, C_2$ are constants.

The admissibility conditions (the negative definiteness of the form $\kappa$) have the form 
\[
\frac{\lambda_3}{\lambda_1-\lambda_4\entr} >0, \quad
\frac{\lambda_5C_1}{2\lambda_4-\lambda_5} >0
\]
for all $\entr \in(-\infty,\entr_0]$.
\begin{theorem}
	The thermodynamic states admitting a one-dimensional symmetry algebra  have the form 
\begin{equation*}
\press =C_1\dens^{\frac{\lambda_5}{2\lambda_4-\lambda_5}}     - \frac{\lambda_2 }{\lambda_5}  ,\quad
\temp =C_2(\lambda_1-\lambda_4\entr)^{-\frac{\lambda_3}{\lambda_4}}   ,
\end{equation*}
	where the constants defining the symmetry algebra satisfy inequalities 
	\[
	\entr_0<\frac{\lambda_1}{\lambda_4}, \quad
	C_1>0, \quad   \frac{\lambda_2 }{\lambda_5}<0, \quad
	\frac{\lambda_5}{2\lambda_4-\lambda_5} >0,
	\]
	and besides they must meet one of the following conditions:
	\begin{enumerate}
		\item if $\frac{\lambda_3}{\lambda_4}$ is irrational, then  $\lambda_3>0$, $\lambda_4>0$, $C_2>0$;
		\item if $\frac{\lambda_3}{\lambda_4}$ is rational, then $\frac{\lambda_3}{\lambda_4}>0$ (i.e. $\frac{\lambda_3}{\lambda_4}=\frac{m}{k}$ ) and
		\begin{enumerate}
			\item if $k$ is even, then $\lambda_4>0$, $C_2>0$;
			\item if $k$ is odd and $m$ is even, then $C_2>0$;
			\item if $k$ is odd and $m$ is odd, then $C_2\lambda_4>0$.
		\end{enumerate}
	\end{enumerate}
\end{theorem}

\subsubsection*{States with a two-dimensional symmetry algebra}

As in the plane case there are no thermodynamic states that admit a two-dimensional non-commutative symmetry algebra.

\subsubsection*{States with a two-dimensional commutative symmetry algebra}

Let now $\LieAlgebra{h_t} \subset \LieAlgebra{h}$ be a commutative two-dimensional Lie subalgebra,
and let $A=\sum\limits_{i=1}^5 \alpha_iY_i$, $B=\sum\limits_{i=1}^5 \beta_iY_i$
be basis vectors in this algebra.

Then the condition $[A,B]=0$ gives the following relations on $\alpha$'s and $\beta$'s:
\begin{equation} \label{commut}
\alpha_1\beta_4 - \alpha_4\beta_1 = 0, \qquad \alpha_2\beta_5 - \alpha_5\beta_2 = 0 .
\end{equation}

Then, as above, restriction of the forms $ \iota_{A}{\Omega}$ and $\iota_{B}{\Omega}$
on the state surface $\lagrangianSurface$ leads us to the four  differential equations of the form \eqref{qwe1}, and
the formal integrability condition for obtained system has the form
\begin{equation*}
{(\alpha_2\beta_5-\alpha_5\beta_2)(5\beta_4-3\beta_5)}=0,
\end{equation*}
which is satisfied due to relations \eqref{commut}.

Solving this system for the general parameters $\alpha$ and $\beta$ we get only the ``nonphysical'' solution of the form $\energy = C_1\dens^{-1} +C_2$.

For the special case, for example, 
\[
 \alpha_3 = \frac{\alpha_4\beta_3}{\beta_4} , \quad
 \alpha_5 = \frac{\alpha_4\beta_5}{\beta_4} 
\]
we get
\[
\press =C_1\dens^{\frac{\beta_5}{2\beta_4-\beta_5}} - \frac{\beta_2}{\beta_5} ,
\quad
\temp =  C_2\left(\entr-\frac{\beta_1}{\beta_4} \right)^{- \frac{\beta_3}{\beta_4}} .
\]

And the admissibility condition leads to the  relations
\[
\frac{\beta_3}{\beta_1-\beta_4\entr} >0, \quad 
\frac{C_1\beta_5}{2\beta_4-\beta_5}>0.
\]
\begin{theorem}
	 In the general case, there are no physically applicable
	thermodynamic states, which admit a two-dimensional commutative symmetry algebra.
	
	For the special case $\alpha_3 = \frac{\alpha_4\beta_3}{\beta_4}$ and $\alpha_5 = \frac{\alpha_4\beta_5}{\beta_4}$ , the thermodynamic states admitting a two-dimensional commutative symmetry algebra have the form 
\[
\press =C_1\dens^{\frac{\beta_5}{2\beta_4-\beta_5}} - \frac{\beta_2}{\beta_5} ,
\quad
\temp =  C_2\left(\entr-\frac{\beta_1}{\beta_4} \right)^{- \frac{\beta_3}{\beta_4}} ,
\]
	where the constants defining the symmetry algebra satisfy inequalities
	\[
	\entr_0<\frac{\beta_1}{\beta_4}, \quad
	C_1>0, \quad   \frac{\beta_2 }{\beta_5}<0, \quad
	\frac{\beta_5}{2\beta_4-\beta_5} >0, \quad
	\frac{\beta_3}{\beta_4}=\frac{m}{k}>0,
	\]
	i.e. $\frac{\beta_3}{\beta_4}$ is rational positive number, and the following cases are possible:
		\begin{enumerate}
			\item if $k$ is odd and $m$ is even, then $C_2>0$;
			\item if $k$ is odd and $m$ is odd, then $C_2\lambda_4>0$.
		\end{enumerate}
\end{theorem}


\subsubsection{Differential invariants}

As in the previous section, we consider two group actions on the Euler equation $\systemEk{}$, i.e. the prolonged  action of the group generated
by the action of the Lie algebra $\LieAlgebra{g_{m}}$ and the action generated by the prolongation of the action of the Lie algebra $\LieAlgebra{g_{sym}}$. So we get two types of differential invariants -- the kinematic and the Euler invariants.

\subsubsection*{The field of kinematic invariants}

First of all, the functions
\[
\dens, \qquad \entr,  \qquad g(\bvec{u},\bvec{u})
\]
(as well as $\press$ and $\temp$) generate all $\LieAlgebra{g_{m}}$-invariants of order zero.

Consider two vector fields $\bvec{u}$ and $\tilde{\bvec{u}}$ such that $g(\bvec{u},\tilde{\bvec{u}})=0$ and $g(\bvec{u},\bvec{u})=g(\tilde{\bvec{u}},\tilde{\bvec{u}})$. Writing the covariant differential  $d_{\nabla}\bvec{u}$ with respect to the vectors $\bvec{u}$ and $\tilde{\bvec{u}}$
as the sum of its symmetric and antisymmetric parts we obtain the 4 invariants of the first order:
\begin{equation}\label{eq:invars1Es}
\begin{aligned}
&J_1 = u_x + v_y + v\cot y, \qquad 
J_2 = u_y\sin y - \frac{v_x}{\sin y} +2 u \cos y, \\
&J_3 = (u(u_xv-v_xu)+v(u_yv-v_yu))\sin y +u \cos y (u^2\sin^2y+2v^2), \\
&J_4 = v(u_xv-v_xu)-u(u_yv-v_yu)\sin^2 y  + v^3\cot y.
\end{aligned}
\end{equation}

The proof of the following theorem can be found in \cite{Duyunova2017-2}.
\begin{theorem} \cite{Duyunova2017-2}
	The following derivations 
	\[
	\nabla_1 = \totalDiff{t} ,\quad
	\nabla_2 = \frac{\dens_x}{\sin^2 y}
	\totalDiff{x}+\dens_y\totalDiff{y},\quad
	\nabla_3 = \frac{\entr_x}{\sin^2 y} \totalDiff{x}+\entr_y\totalDiff{y}
	\]
	are $\LieAlgebra{g_{m}}$-invariant. They are linearly independent if 
	\[
	\dens_x\entr_y-\dens_y\entr_x\neq 0.
	\]
\end{theorem}

It is easy to check that the codimension of regular $\LieAlgebra{g_{m}}$-orbits is equal to 12. The Rosenlicht theorem \cite{Ros} gives us the following result.
\begin{theorem} \cite{Duyunova2017-2}
	The field of the first order kinematic invariants 
	is generated by the invariants $\dens,\, \entr,\, g(\bvec{u},\bvec{u})$ of order zero and by the invariants \eqref{eq:invars1Es} and 
	\begin{equation}\label{eq:invars2Es}
	\nabla_1\dens, \quad \nabla_1\entr, \quad 
	\nabla_2\dens, \quad  \nabla_2\entr, \quad \nabla_3\entr
	\end{equation} 
	 of order one. 
	These invariants separate regular $\LieAlgebra{g_m}$-orbits.
\end{theorem}

The bundle $\pi_{2,1}:\systemEk{2}\rightarrow \systemEk{1}$ has rank 14,
and by applying the derivations $\nabla_1, \nabla_2, \nabla_3$ to the kinematic invariants \eqref{eq:invars1Es} and \eqref{eq:invars2Es} we get 27 kinematic invariants. Straightforward computations show that among these invariants 14 are always independent (see  \textit{http://d-omega.org}).

Therefore, starting with the order $k=1$ dimensions of regular orbits are equal to $\dim \LieAlgebra{g_m}=4$.

The Hilbert function (the number of independent invariants) of the $\LieAlgebra{g_m}$-invariants field has form $H(k) = 5k +4$ for $k\geq 1$ and $H(0)=3$, and 
the corresponding Poincar\'{e} function is equal to
\begin{equation*}
P(z) = \frac{3+3z-z^3}{(1-z)^2}.
\end{equation*}

Summarizing, we get the following result.
\begin{theorem} \cite{Duyunova2017-2}
	The field of the kinematic invariants is generated by the invariants $\dens, \,  \entr, \, g(\bvec{u},\bvec{u})$ of order zero, by the invariants \eqref{eq:invars1Es} and \eqref{eq:invars2Es} of order one and by the invariant derivations  $\nabla_1, \nabla_2, \nabla_3$.
	This field separates regular orbits.
\end{theorem}

\subsubsection*{The field of Euler invariants}

Let us consider the case when the equations of thermodynamic state  $\lagrangianSurface$  admit a one-dimen\-sio\-nal symmetry algebra
generated by the vector field

\begin{equation*}
A =  \xi_1 X_5 + \xi_2 X_6 + \xi_3 X_7 + \xi_4 X_{8} + \xi_5 X_{9} .
\end{equation*}

Using a similar computations as in the plane case we get the following result.
\begin{theorem} \cite{Duyunova2017-2}
	The field of the Euler differential invariants on a sphere for thermodynamic states
	admitting a one-dimensional symmetry algebra is generated by
	the differential invariants 
	\[
	J_1 \dens \left( \entr - \frac{\xi_1}{\xi_4+\xi_5}  \right) , \quad 
	J_1 \dens^{\frac{\xi_4}{2\xi_4+\xi_5}}, \quad
	\frac{g(\bvec{u},\bvec{u})}{J_1^2}, \\
	\]
	\[
	\frac{J_2}{J_1}, \quad
	\frac{J_3}{J_1^3},  \quad
	\frac{J_4}{J_1^3}, \quad 
	\frac{\nabla_1\dens}{ J_1\dens}, \quad
	\frac{\nabla_2\dens}{\dens^2}, \quad 
	\dens\nabla_1\entr, \quad
	J_1\nabla_2\entr, \quad
	J_1^2\dens^2\nabla_3\entr
	\]
	of the first order and by the invariant derivations 
	\begin{equation*} 
	\dens^{\frac{\xi_4}{2\xi_4+\xi_5}} \nabla_1,\quad
	\dens^{-1} \nabla_2,\quad
	\dens^{\frac{\xi_4+\xi_5}{2\xi_4+\xi_5}} \nabla_3 .
	\end{equation*}
	This field separates regular orbits.
\end{theorem}

The last formulas are valid for general $\xi$'s. All details and the special cases are considered in \cite{Duyunova2017-2}.

Now let the thermodynamic state admit a commutative two-dimensional symmetry algebra generated by the vector fields 
$A=\sum\limits_{i=1}^5 \mu_iX_{i+4}$, $B=\sum\limits_{i=1}^5 \eta_iX_{i+4}$ such that $\mu$'s and $\eta$'s satisfy relations 
\begin{equation*}
\left\{
\begin{aligned}
& \eta_1\mu_4-\eta_4\mu_1+\eta_1\mu_5-\eta_5\mu_1=0, \\
& \eta_2\mu_5-\eta_5\mu_2=0.
\end{aligned} \right. 
\end{equation*}
\begin{theorem} \cite{Duyunova2017-2}
	The field of Euler differential invariants for the thermodynamic states  admitting
	a com\-mu\-ta\-tive two-dimensional symmetry algebra is generated by differential invariants 
	\[
	J_1 \dens ((\mu_4+\mu_5)\entr -\mu_1), \quad
	\frac{g(\bvec{u},\bvec{u})}{J_1^2}, \quad 
	\frac{J_2}{J_1}, \quad
	\frac{J_3}{J_1^3},\quad
	\frac{J_4}{J_1^3},  
	\]
	\[
	\frac{\nabla_1\dens}{ J_1\dens}, \quad
	\dens\nabla_1\entr, \quad
	\frac{\nabla_2\dens}{\dens^2}, \quad 
	J_1\nabla_2\entr, \quad
	J_1^2\dens^2\nabla_3\entr
	\]
	of the first order and by the invariant derivations 
	\[
	\dens ((\mu_4+\mu_5)\entr -\mu_1) \nabla_1,\quad
	\dens^{-1} \nabla_2,\quad
	((\mu_4+\mu_5)\entr -\mu_1)^{-1} \nabla_3 .
	\]
	This field separates regular orbits.
\end{theorem}

\subsection{Flows on a spherical layer}
Consider Euler system \eqref{eq:E} on a spherical layer $M=S^2\times \mathbb{R}$ with the coordinates $(x,y,z)$, where $(x,y)$ are the stereographic coordinates on the sphere, and the metric 
\[
g=\frac{4}{(x^2+y^2+1)^2}(dx^2+dy^2) +dz^2.
\]

The velocity field of the flow has the form 
$\bvec{u}=u(t,x,y,z)\,\dir{x} +v(t,x,y,z)\,\dir{y}+w(t,x,y,z)\,\dir{z}$, 
the pressure $\press$, the density $\dens$, the temperature $\temp$ and the entropy $\entr$ are the functions of time and space with the coordinates $(t,x,y,z)$.

The vector of gravitational acceleration is of the form $\bvec{g}={(0,0,g)}$.

\subsubsection{Symmetry Lie algebra}

Consider the Lie algebra $\LieAlgebra{g}$ generated
by the following vector fields on the manifold $\JetSpace{0}{\pi}$: 
\begin{equation*} 
\begin{aligned} 
&X_1 = \dir{ t}, \qquad
X_3 = t\,\dir{ z}+\dir{ w}, \qquad\\
&X_2 = \dir{ z}, \qquad 
X_4 = y\,\dir{ x}-x\,\dir{y}+v\,\dir{u}-u\,\dir{v}, \\
&X_{5} = xy \,\dir{ x} -\frac{1}{2} (x^2-y^2-1)\dir{ y} +
\left( xv + yu \right) \dir{ u} - (xu-yv)\dir{ v}  ,  \\
&X_{6} =    \frac{1}{2}(x^2-y^2+1)\dir{ x} + xy \,\dir{ y} +
(xu-yv) \dir{ u} + (xv+yu)\dir{ v}  ,  \\
&X_{7} =  \dir{ \entr} ,  \qquad
X_{10} = t\,\dir{ t} +\mathrm{g}t^2\,\dir{ z} - u\,\dir{ u} - v\,\dir{ v} +(2\mathrm{g}t- w)\dir{ w} + 2\dens\,\dir{ \dens} - \entr\,\dir{ \entr},\\
&X_{8} =  \dir{ \press} , \qquad
X_{11} = \press\,\dir{ \press} + \dens\,\dir{ \dens} - \entr\,\dir{ \entr} , \\
&X_{9} = \temp\,\dir{ \temp}  .
\end{aligned} 
\end{equation*}

The pure geometric part $\LieAlgebra{g_{m}}$ generated by the vector fields $X_1,X_2, \ldots, X_6$. Transformations corresponding to the elements of the Lie group generated by the algebra $\LieAlgebra{g_{m}}$ are compositions of sphere motions, Galilean transformations and shifts along the $z$ direction, time shifts. 

To describe thermodynamic part of the symmetry algebra, we consider the Lie algebra $\LieAlgebra{h}$ generated by the vector fields
\begin{equation*}
\begin{aligned}
&Y_1 = \dir{ \entr}, \qquad
Y_2 = \dir{ \press}, \qquad
Y_3 =  \temp\,\dir{ \temp} ,\\
&Y_4 =  2\dens\,\dir{ \dens} - \entr\,\dir{ \entr}, \qquad
Y_{5} = \press\,\dir{ \press} - \dens\,\dir{ \dens}  .
\end{aligned}
\end{equation*}

This is a solvable Lie algebra with the following structure
\[
\left[ Y_1,Y_4\right] =-Y_1, \qquad  \left[ Y_2,Y_5\right] =Y_2.
\]

Let also $\LieAlgebra{h_{t}}$ be the Lie subalgebra of algebra $\LieAlgebra{h}$ that preserves thermodynamic state \eqref{eq:Therm}.
Then the following result is valid.
\begin{theorem} \cite{Duyunova2017-3}
	The Lie algebra $\LieAlgebra{g_{sym}}$ of point symmetries of the Euler system of differential equations on a spherical layer coincides with 
	\[
	\vartheta^{-1}(\LieAlgebra{h_{t}}).
	\]
\end{theorem}

\subsubsection{Symmetry classification of states}
	
The Lie algebra generated by the vector fields $Y_1, \ldots, Y_5$ coincides with the Lie algebra of the thermodynamic symmetries of the Euler system on a sphere. 

Thus the classification of the thermodynamic states or Lagrangian surfaces  $\lagrangianSurface$ depending on the dimension of the symmetry algebra $\LieAlgebra{h_{t}}\subset \LieAlgebra{h}$ is the same as the classification presented in the previous section.


\subsubsection{Differential invariants}

\subsubsection*{The field of kinematic invariants}

First of all,  the functions
$\dens$, $\entr$, $g(\bvec{u},\bvec{u})-w^2$ 
(as well as $\press$ and $\temp$) generate all $\LieAlgebra{g_{m}}$-invariants of order zero.

The proofs of the following theorems can be found in \cite{Duyunova2017-3}.
\begin{theorem} \cite{Duyunova2017-3}
	The following derivations 
	\[
	\nabla_1 = \totalDiff{z} ,\qquad
	\nabla_2 = \totalDiff{t}+w\totalDiff{z},\qquad
	\nabla_3 = u\totalDiff{x}+v\totalDiff{y},\qquad
	\nabla_4 =  v \totalDiff{x}-u\totalDiff{y}  
	\]	
	are $\LieAlgebra{g_{m}}$-invariant. They are linearly independent if 
	\[
	u^2+v^2\neq 0.
	\]
\end{theorem}
\begin{theorem} \cite{Duyunova2017-3}
	The field of the first order kinematic invariants 
	is generated by the invariants $\dens,\, \entr,\, g(\bvec{u},\bvec{u})-w^2$ of order zero and by the invariants 
	\begin{equation}\label{jkh}
	\begin{aligned}
	&\nabla_1\dens, \quad \nabla_2\dens, \quad \nabla_3\dens, \quad \nabla_4\dens, \quad
	\nabla_1\entr, \quad \nabla_2\entr, \quad \nabla_3\entr, \quad \nabla_4\entr,	\\
	&\nabla_1(g(\bvec{u},\bvec{u})-w^2), \quad \nabla_3(g(\bvec{u},\bvec{u})-w^2), \quad \nabla_4(g(\bvec{u},\bvec{u})-w^2),\\
	 &\nabla_1w, \quad  \nabla_3w, \quad \nabla_4w, \quad
	 J_1 = u_z w_x+v_z w_y, \quad J_2=\frac{u_t v_z-u_z v_t}{u_z^2+v_z^2}
	 \end{aligned}
	 \end{equation}
	of order one. 
	These invariants separate regular $\LieAlgebra{g_m}$-orbits.
\end{theorem}

The bundle $\pi_{2,1}:\systemEk{2}\rightarrow \systemEk{1}$ has rank 33,
and by applying the derivations $\nabla_i$, $i=1,\ldots,4$ to the first order kinematic invariants \eqref{jkh} we get 64 kinematic invariants.
Straightforward computations show that among these invariants 33 are always independent.

Therefore, starting with the order $k=1$ dimensions of the  regular orbits are equal to $\dim \LieAlgebra{g_m}=6$.

Moreover, the number of independent invariants (the Hilbert function) is equal to $H(k) = 3k^2+ 8k +5$ for $k\geq 1$ and $H(0)=3$.

The corresponding Poincar\'{e} function has the form
\begin{equation*}
P(z) = \frac{3+7z-6z^2+2z^3}{(1-z)^3}.
\end{equation*}
\begin{theorem} \cite{Duyunova2017-3}
	The field of the kinematic invariants is generated by the invariants $\dens, \,  \entr, \, g(\bvec{u},\bvec{u})-w^2$ of order zero,
	by the invariants \eqref{jkh} of order one and by the invariant derivations  $\nabla_i$, $i=1,\ldots,4$.
	This field separates regular orbits.
\end{theorem}

\subsubsection*{The field of Euler invariants}

At first we consider the case when the thermodynamic state  $\lagrangianSurface$ admits a one-dimensional symmetry algebra generated by the vector field
\begin{equation*}
A =  \xi_1 X_7 + \xi_2 X_8 + \xi_3 X_9 + \xi_4 X_{10} + \xi_5 X_{11} .
\end{equation*}

Then for general values of the parameters $\xi$'s we have the following result. The special cases are considered in \cite{Duyunova2017-3}.
\begin{theorem} \cite{Duyunova2017-3}
	The field of the Euler differential invariants for thermodynamic states
	admitting a one-dimensional symmetry algebra is generated by
	the differential invariants 
	\begin{equation*}
	\begin{aligned}
	& w_z \dens \left( \entr - \frac{\xi_1}{\xi_4+\xi_5}  \right) , \quad 
	w_z^{-2}\left( {g(\bvec{u},\bvec{u})-w^2}\right) , \quad \\
	&\frac{\nabla_1\dens}{\dens}, \quad 
	\frac{\nabla_2\dens}{w_z \dens}, \quad
	\frac{\nabla_3\dens}{w_z \dens}, \quad
	\frac{\nabla_4\dens}{w_z \dens},  \quad \\
	&w_z\dens\nabla_1\entr, \quad 
	\dens\nabla_2\entr, \quad 
	\dens\nabla_3\entr, \quad
	\dens\nabla_4\entr, \quad  \\
	&w_z^{-2}{\nabla_1\left( g(\bvec{u},\bvec{u})-w^2\right) }, \quad 
	w_z^{-3}{\nabla_3\left( g(\bvec{u},\bvec{u})-w^2\right) }, \quad 
	w_z^{-3}{\nabla_4\left( g(\bvec{u},\bvec{u})-w^2\right) }, \quad \\
	&w_z \dens^{\frac{\xi_4}{2\xi_4+\xi_5}}, \quad 
	w_z^{-2}{\nabla_3 w}, \quad
	w_z^{-2}{\nabla_4 w}, \quad
	w_z^{-2}{J_1}, \quad
	w_z^{-1}{J_2},
	\end{aligned}
	\end{equation*}
	of the first order and by the invariant derivatives 
	\begin{equation*} 
	\nabla_1,\quad
	w_z^{-1} \nabla_2,\quad
    w_z^{-1} \nabla_3,\quad
	w_z^{-1} \nabla_4 .
	\end{equation*}
	This field separates regular orbits.
\end{theorem}

Now, let the thermodynamic state admit a commutative two-di\-men\-sional symmetry algebra generated by the vector fields 
$A=\sum\limits_{i=1}^5 \mu_iX_{i+6}$, $B=\sum\limits_{i=1}^5 \eta_iX_{i+6}$, then $\mu$'s and $\eta$'s satisfy relations 
\begin{equation*}
\left\{
\begin{aligned}
& \eta_1\mu_4-\eta_4\mu_1+\eta_1\mu_5-\eta_5\mu_1=0, \\
& \eta_2\mu_5-\eta_5\mu_2=0.
\end{aligned} \right. 
\end{equation*}
\begin{theorem} \cite{Duyunova2017-3}
	The field of Euler differential invariants for thermodynamic states admitting
	a com\-mu\-ta\-tive two-dimensional symmetry algebra is generated by differential invariants 
	\[
	\begin{aligned}
	&
	w_z^{-2}\left( {g(\bvec{u},\bvec{u})-w^2}\right) , \quad \\
	&\frac{\nabla_1\dens}{\dens}, \quad 
	\frac{\nabla_2\dens}{w_z \dens}, \quad
	\frac{\nabla_3\dens}{w_z \dens}, \quad
	\frac{\nabla_4\dens}{w_z \dens}, \quad \\
	&w_z\dens\nabla_1\entr, \quad 
	\dens\nabla_2\entr, \quad 
	\dens\nabla_3\entr, \quad
	\dens\nabla_4\entr, \quad  \\
	&w_z^{-2}{\nabla_1\left( g(\bvec{u},\bvec{u})-w^2\right) }, \quad 
	w_z^{-3}{\nabla_3\left( g(\bvec{u},\bvec{u})-w^2\right) }, \quad 
	w_z^{-3}{\nabla_4\left( g(\bvec{u},\bvec{u})-w^2\right) }, \quad \\
	&w_z \dens ((\mu_4+\mu_5)\entr -\mu_1), \quad
	w_z^{-2}{\nabla_3 w}, \quad
	w_z^{-2}{\nabla_4 w}, \quad
	w_z^{-2}{J_1}, \quad
	w_z^{-1}{J_2}
	\end{aligned}
	\]
	of the first order and by the invariant derivatives 
	\[
	\nabla_1,\quad
	w_z^{-1} \nabla_2,\quad
	w_z^{-1} \nabla_3,\quad
	w_z^{-1} \nabla_4 .
	\]
	This field separates regular orbits.
\end{theorem}

\newpage
\section{Compressible viscid fluids or gases}\label{sec:viscid}

In this section we study differential invariants of compressible viscid fluids or gases. 

The system of differential equations (the Navier--Stokes system) describing flows on an oriented Riemannian manifold $(M, g)$ consists of the following equations (see \cite{Batchelor2000} for details):
\begin{equation}\label{eq:NS}
\left\{
\begin{aligned}
& \dens(\bvec{u}_t  + \nabla_{\bvec{u}}\bvec{u})- \diver\stress -\bvec{g}\dens =0  ,\\
& \parder{(\dens\, \Omega_g)}{t} + \LieDerivative{\bvec{u}}{\dens\, \Omega_g} = 0,\\
&\dens\temp\left(  \entr_t + \nabla_{\bvec{u}}s\right) - \Phi + k (\Delta_g T) =0.
\end{aligned}
\right.
\end{equation}

Here the divergence operator $\diver : S^2T^*M \rightarrow TM$ is given by
\[
\left( \diver \sigma\right) _l = (d_{\nabla}\sigma)_{ijk}g^{jk}g^{il},
\]
where $d_{\nabla}$ is the covariant differential.

The fluid under consideration is assumed to be newtonian and isotropic.
Therefore, the fluid stress tensor $\stress$ is symmetric, and it depends
on the rate of deformation tensor
$\Def=\frac{1}{2}\LieDerivative{\bvec{u}}{g}$ linearly. These two conditions
give the following form of the stress tensor: $\stress=-\press g + \visc$, where the viscous
stress tensor $\visc$ is given by
\[
\visc = 2\eta\left(\Def- \frac{\inner{\Def}{g}}{\inner{g}{g}}g \right)+\zeta\inner{\Def}{g}g.
\]

The quantity $\Phi=\inner{\visc}{\Def}$ represents the rate of dissipation
of mechanical energy~\cite{Batchelor2000}.

The first equation of system~\eqref{eq:NS} is the 
Navier--Stokes equation, the second one is the continuity equation and the third one is the general equation of heat transfer.

In this section we consider the following examples of manifold $M$: a plane, a three-dimensional space, a sphere and a  spherical layer. 

Note that in all these cases the number of unknown functions is greater than the number of system equations by 2, i.e. the system \eqref{eq:NS} is incomplete. As above we get two additional equations
using the thermodynamics of the medium.

Thus, by the Navier--Stokes system of differential equations we mean the system~\eqref{eq:NS}
extended by two equations of state~\eqref{eq:Therm},
where functions $f$ and $g$ satisfy the additional relation~\eqref{eq:lagr} and the form $\kappa$ is negative definite.

Geometrically, we represent this system in the following way.
Consider the bundle 
\[
\pi : \mathbb{R} \times TM \times \mathbb{R}^4  
\longrightarrow \mathbb{R}\times M
\]
of rank $(\dim  M+4)$. 

Then the Navier--Stokes system is a system
of differential equations on sections of the bundle $\pi$.

Note that system~\eqref{eq:Therm} defines the zeroth
order system $\systemEk{0}\subset \JetSpace{0}{\pi}$. 

Denote by $\systemEk{1}\subset \JetSpace{1}{\pi}$ the system of order $\leq 1$ obtained by the first prolongation of the system $\systemEk{0}$
and by the continuity equation of system~\eqref{eq:NS}. 

Let also $\systemEk{2}\subset \JetSpace{2}{\pi}$ be the system of differential
equations of order $\leq 2$ obtained by the first prolongation
of the system $\systemEk{1}$ and all equations of system~\eqref{eq:NS}.

For the case $k\geq 3$, we define $\systemEk{k}\subset\JetSpace{k}{\pi}$
to be the $(k-2)$-th prolongation of the system $\systemEk{2}$.

Note that the system $\systemEk{\infty}=\lim\limits_{\longleftarrow}\systemEk{k}$ is a formally
integrable system of differential equations, which we also call the Navier--Stokes system.


\subsection{2D-flows}

Consider Navier--Stokes system \eqref{eq:NS} on a plane $M=\mathbb{R}^2$ equipped  with the coordinates $(x,y)$ and the standard flat metric $g=dx^2+dy^2$. 

The velocity field of the flow has the form 
$\bvec{u}=u(t,x,y)\,\dir{x} +v(t,x,y)\,\dir{y}$, 
the pressure $\press$, the density $\dens$, the temperature $\temp$ and the entropy $\entr$ are the functions of time and space with the coordinates $(t,x,y)$.

Here we also consider the flow without any external force field, so $\bvec{g}={0}$.


\subsubsection{Symmetry Lie algebra}

To describe the Lie algebra of symmetries of the Navier--Stokes system we consider a Lie algebra $\LieAlgebra{g}$ generated by the following vector fields on space  $\JetSpace{0}{\pi}$:
\begin{equation*}
\begin{aligned} 
&X_1 = \dir{ x}, \qquad  
\phantom{y\,- x\,\dir{ y} + v\,\dir{ u} - u\,\dir{ v}  }
X_4 = t\,\dir{ x} + \dir{ u}, \qquad\\
&X_2 = \dir{ y}, \qquad 
\phantom{y\,- x\,\dir{ y} + v\,\dir{ u} - u\,\dir{ v}  }
X_5 = t\,\dir{ y} + \dir{ v} , \\
&X_{3} =  y\,\dir{ x} - x\,\dir{ y} + v\,\dir{ u} - u\,\dir{ v}  , 
 \qquad X_6 = \dir{t},  \\[2pt]
&X_7 = \dir{ \entr}, \qquad
X_8 = \dir{ \press} , \qquad \\
&X_{9} =  x\,\dir{ x} + y\,\dir{ y} + u\,\dir{ u} + v\,\dir{ v} - 2\dens\,\dir{ \dens} +2\temp\,\dir{ \temp} , \\
&X_{10} = t\,\dir{ t} - u\,\dir{ u} - v\,\dir{ v} + \dens\,\dir{ \dens} - \press\,\dir{ \press} -2 \temp\,\dir{ \temp} .
\end{aligned} 
\end{equation*}

In general the symmetry algebra of system \eqref{eq:NS} consists of pure geometric and thermodynamic parts. 

The  geometric part is represented 
by the algebra $\LieAlgebra{g_{m}}=\left\langle X_1,X_2, \ldots, X_6\right\rangle  $ with respect to the group of motions, Galilean transformations and time shifts. 

Moreover, the kernel of homomorphism $\vartheta$ \eqref{theta} is an ideal  $\LieAlgebra{g_{m}}$ in the Lie algebra $\LieAlgebra{g}$.

The thermodynamic part strongly depends on the symmetries of the thermodynamic state. In order to describe it, denote by $\LieAlgebra{h}$ a Lie algebra generated by the vector fields
\begin{equation*}
Y_1=\dir{\entr} ,  \quad    Y_2=\dir{\press}, \quad 
Y_3=\dens\,\dir{\dens}-\temp\,\dir{\temp},  \quad    
Y_4=\press\,\dir{\press}+\temp\,\dir{\temp}.
\end{equation*}

Let also $\LieAlgebra{h_{t}}$ be a Lie subalgebra of the algebra $\LieAlgebra{h}$ which preserves the thermodynamic state \eqref{eq:Therm}.
\begin{theorem} \cite{Duyunova2017-4}
	A Lie algebra $\LieAlgebra{g_{sym}}$ of symmetries of the Navier--Stokes system of differential equations on a plane coincides with 
	\[
	\vartheta^{-1}(\LieAlgebra{h_{t}}).
	\]
\end{theorem}

Note that, usually, the equations of state are neglected and the vector fields like $f(t)\,\dir{ \press}$, where $f$ is an arbitrary function, considered as symmetries of the Navier--Stokes system.

For the general equation of state $\LieAlgebra{h_{t}}=0$ and the symmetry algebra coincides with the algebra $\LieAlgebra{g_{m}}$.

\subsubsection{Symmetry classification of states}

In this section we classify thermodynamic states or Lagrangian surfaces  $\lagrangianSurface$ depending on the dimension of the symmetry algebra $\LieAlgebra{h_{t}}\subset \LieAlgebra{h}$.

We consider one- and two-dimensional symmetry algebras only. One can easily check that there are no physically valuable thermodynamic states with three or more dimensional symmetry algebras.

\subsubsection*{States with a one-dimensional symmetry algebra}

Let $\dim \LieAlgebra{h_{t}}=1 $ and let $Z = \sum\limits_{i=1}^4 \lambda_i Y_i$ be a basis vector in this algebra, then the differential 1-form $\iota_{Z}{\Omega}$ has the form
$$
\iota_{Z}{\Omega} = -\frac{\lambda_3}{\dens}\,d\press
+\frac{\lambda_4\press+\lambda_2}{\dens^2}\,d\dens
+(\lambda_3-\lambda_4)\temp d\entr-\lambda_1 d\temp,
$$
and, in terms of specific energy $\energy(\dens,\entr)$, the Lagrangian surface $\lagrangianSurface$ can be found as a solution of the following PDE system
\begin{equation*}
\left\lbrace
\begin{aligned}
&\lambda_1{\energy}_{\entr\entr}+\lambda_3\dens{\energy}_{\entr\dens}+
(\lambda_3-\lambda_4){\energy}_{\entr}=0,\\
&\lambda_3\dens{\energy}_{\dens\dens}+
\lambda_1{\energy}_{\entr\dens}+(2\lambda_3-\lambda_4){\energy}_{\dens}
-\frac{\lambda_2}{\dens^2}=0.\\
\end{aligned}
\right.
\end{equation*}

It is easy to check that the bracket of these two equations (see \cite{Kruglikov2002}) vanishes and therefore the system is formally integrable and compatible.

Solving this system for general values of parameters $\lambda$, all special cases are considered in \cite{Duyunova2017-4}, we get expressions for the presser and the temperature
\[
\temp=\dens^{\frac{\lambda_4}{\lambda_3}-1}F^{\prime},\quad
\press=\dens^{\frac{\lambda_4}{\lambda_3}}\left(
\left(\frac{\lambda_4}{\lambda_3}-1\right)F-\frac{\lambda_1}{\lambda_3}
F^{\prime}\right)-\frac{\lambda_2}{\lambda_4},
\quad
F= F\left(\entr-\frac{\lambda_1}{\lambda_3}\ln\dens\right),
\]
where $F$
is a smooth function.

Negative definiteness of the quadratic form $\kappa$ gives the following relations on the function $F$ and the parameters $\lambda$: 
\[
\lambda_1^2 \dens^{\frac{\lambda_4}{\lambda_3}} F^{\prime\prime} +
\lambda_1(\lambda_3-\lambda_4)\dens\temp + \lambda_3(\lambda_4\press +\lambda_2)>0 ,
\]
\[
\dens^{\frac{\lambda_4-2\lambda_3}{\lambda_3}} F^{\prime\prime}
 \left( \lambda_1(\lambda_3-\lambda_4)\dens\temp - \lambda_3(\lambda_4\press +\lambda_2)\right) +\temp^2(\lambda_3-\lambda_4)^2<0.
\]

\begin{theorem}
	The thermodynamic states admitting a one-dimensional symmetry algebra have the form 
\[
\temp=\dens^{\frac{\lambda_4}{\lambda_3}-1}F^{\prime},\quad
\press=\dens^{\frac{\lambda_4}{\lambda_3}}\left(
\left(\frac{\lambda_4}{\lambda_3}-1\right)F-\frac{\lambda_1}{\lambda_3}
F^{\prime}\right)-\frac{\lambda_2}{\lambda_4},
\quad
F= F\left(\entr-\frac{\lambda_1}{\lambda_3}\ln\dens\right),
\]
	where $F$ is a smooth function, $F^{\prime}$ is positive and
	\[
	\lambda_1^2  F^{\prime\prime} +	\lambda_1(\lambda_3-2\lambda_4) F^{\prime} + \lambda_4(\lambda_4-\lambda_3)F>0, \quad
	\]
	\[
	 F^{\prime\prime} (\lambda_4(\lambda_4-\lambda_3)F -\lambda_1\lambda_3F^{\prime} ) - (F^{\prime} )^2(\lambda_4-\lambda_3)^2>0.
	\]
\end{theorem}

\subsubsection*{States with a two-dimensional non-commutative symmetry algebra}

Let $\LieAlgebra{h_t} \subset \LieAlgebra{h}$ be a non-commutative two-dimensional Lie subalgebra. It is easy to check that two vectors of the form $A=Y_2$, $B=\alpha Y_1 + \beta Y_3 + Y_4$ are the basis vectors in the non-commutative algebra $\LieAlgebra{h_t}$.

Then, as above, the restrictions of forms $ \iota_{A}{\Omega}$ and $\iota_{B}{\Omega}$ on the state surface $\lagrangianSurface$ lead us to the solution $\dens= const$. 

Since we consider thermodynamic states  such that the variables $\dens$ and $\entr$ are local coordinates then we do not consider the case of the non-commutative subalgebra.

\subsubsection*{States with a two-dimensional commutative symmetry algebra}

Let now $\LieAlgebra{h_t} \subset \LieAlgebra{h}$ be a commutative two-dimensional Lie subalgebra, and let $A=\sum\limits_{i=1}^4 \alpha_iY_i$, $B=\sum\limits_{i=1}^4 \beta_iY_i$ be the basis vectors in the algebra $\LieAlgebra{h_t}$.

Then condition $[A,B]=0$ gives the following relations on $\alpha$'s and $\beta$'s:
\begin{equation} \label{commutNS2}
\alpha_2\beta_4 - \alpha_4\beta_2 = 0.
\end{equation}

Then the restrictions of forms $ \iota_{A}{\Omega}$ and $\iota_{B}{\Omega}$ on the state surface $\lagrangianSurface$ lead us to the following system of differential equations:
\begin{equation*}
\left\lbrace
\begin{aligned}
&\alpha_1{\energy}_{\entr\entr}+\alpha_3\dens{\energy}_{\entr\dens}+
(\alpha_3-\alpha_4){\energy}_{\entr}=0,\\
&\alpha_3\dens{\energy}_{\dens\dens}+
\alpha_1{\energy}_{\entr\dens}+(2\alpha_3-\alpha_4){\energy}_{\dens}
-\frac{\alpha_2}{\dens^2}=0,\\
&\beta_1{\energy}_{\entr\entr}+\beta_3\dens{\energy}_{\entr\dens}+
(\beta_3-\beta_4){\energy}_{\entr}=0,\\
&\beta_3\dens{\energy}_{\dens\dens}+
\beta_1{\energy}_{\entr\dens}+(2\beta_3-\beta_4){\energy}_{\dens}
-\frac{\beta_2}{\dens^2}=0.\\
\end{aligned}
\right.
\end{equation*}

The formal integrability condition for this system has the form
\[
{(5\beta_3-\beta_4)(\alpha_2\beta_4-\beta_2\alpha_4)}=0,
\]
which is satisfied due to relations \eqref{commutNS2}.

Solving this PDE system we get the following expressions for the pressure and the temperature
\begin{equation}\label{eq:energyEquation2Sol} 
\press = C(\beta-1)e^{\alpha\entr }\dens^{\beta} - \frac{\beta_2}{\beta_4}, \quad
\temp = C\alpha e^{\alpha\entr }\dens^{\beta-1},
\end{equation}
where 
\[
\alpha = \frac{\alpha_4\beta_3-\alpha_3\beta_4}{\alpha_1\beta_3-\alpha_3\beta_1}, \quad
\beta = \frac{\alpha_1\beta_4-\beta_1\alpha_4}{\alpha_1\beta_3-\alpha_3\beta_1},
\]
and the admissibility conditions have the form $\alpha>0$, $\beta>1$.
\begin{theorem}
	The thermodynamic states admitting a two-dimensional commutative symmetry algebra  have the form 
	\[
\press = C(\beta-1)e^{\alpha\entr }\dens^{\beta} - \frac{\beta_2}{\beta_4}, \quad
\temp = C\alpha e^{\alpha\entr }\dens^{\beta-1},
	\]
	where 
	\[
	\alpha = \frac{\alpha_4\beta_3-\alpha_3\beta_4}{\alpha_1\beta_3-\alpha_3\beta_1}>0, \quad
	\beta = \frac{\alpha_1\beta_4-\beta_1\alpha_4}{\alpha_1\beta_3-\alpha_3\beta_1}>1, \quad
	C>0, \quad   \frac{\beta_2 }{\beta_4}<0. \quad
	\]
\end{theorem}

Observe that, the expressions for the temperature and the pressure for an ideal gas
\[
	\temp = \frac{1}{\gamma}\dens^k e^{\frac{\entr}{\gamma}}, \quad 
	\press = k\dens^{k+1}e^{\frac{\entr}{\gamma}},
\]
where $k$ and $\gamma$ are constant depending on a gas, can be obtained from the equations \eqref{eq:energyEquation2Sol} 
by choosing appropriate values of the constants.

\subsubsection{Differential invariants}

As in the case of compressible inviscid fluids or gases (the Euler system), we consider two group actions on the Navier--Stokes equation $\systemEk{}$. 

The first is the prolonged  action  of the group generated  by the action of Lie algebra $\LieAlgebra{g_{m}}$ and the differential invariants with respect this action we call \textit{kinematic differential invariants}.   

The second action is the action generated by prolongation of the action Lie algebra $\LieAlgebra{g_{sym}}$, and the differential invariants with respect second action we call \textit{Navier--Stokes invariants}.

Also we say that a point $x_k\in \systemEk{k}$ and the corresponding orbit $\mathcal{O}(x_k)$ ($\LieAlgebra{g_{m}}$ or $\LieAlgebra{g_{sym}}$-orbit) are \textit{regular}, if there are exactly $m=\mathrm{codim}\, \mathcal{O}(x_k) $ independent  invariants (kinematic or Navier--Stokes) in a neighbourhood of this orbit. 

Thus, the corresponding point on the quotient space $\systemEk{k} /{\LieAlgebra{g_{m}}}$ or $\systemEk{k} /{\LieAlgebra{g_{sym}}}$ 
is smooth, and these independent invariants  (kinematic or Navier--Stokes) can serve as local coordinates in a neighbourhood of this point.

Otherwise, we say that the point and the corresponding orbit are \textit{singular}. 

It is worth to note that the Navier--Stokes system together with the symmetry algebras
$ \LieAlgebra{g_{m}}$ or $\LieAlgebra{g_{sym}}$ satisfies the conditions
of the Lie--Tresse theorem (see \cite{KL}), and therefore
the above differential invariants separate
regular $ \LieAlgebra{g_{m}}$ or $\LieAlgebra{g_{sym}}$
orbits on the Navier--Stokes system $\systemEk{}$.

\subsubsection*{The field of kinematic invariants}

First of all observe that the density $\dens$ and the entropy $\entr$
(as well as the pressure $\press$ and the temperature $\temp$) on the equation $\systemEk{0}$ are $\LieAlgebra{g_{m}}$-invariants. 

Moreover, the following functions are the kinematic invariants of the first order (see \cite{Duyunova2017-4}):
\begin{equation*}
\begin{aligned}
&J_1=u_x+v_y, \qquad \,J_5= \dens_x\entr_y-\dens_y\entr_x ,\\
&J_2=u_y-v_x  ,\qquad  \,J_6= \entr_t + \entr_x u + \entr_y v,  \\
&J_3 = \dens_x^2 + \dens_y^2, \qquad 
J_7=  \dens_x(\dens_xu_x+\dens_yu_y) + \dens_y(\dens_xv_x+\dens_yv_y), \\
&J_4 = \entr_x^2 + \entr_y^2, \qquad  \,\,
J_{8}= \entr_x(\dens_xu_x+\dens_yu_y) + \entr_y(\dens_xv_x+\dens_yv_y) ,\\
&J_{9}=\entr_x(u_t+uu_x+vu_y)+\entr_y(v_t+uv_x+vv_y),\\
&J_{10}=\dens_x(u_t+uu_x+vu_y)+\dens_y(v_t+uv_x+vv_y).
\end{aligned}
\end{equation*}
\begin{proposition}
	The singular points belong to the union of two sets: 
	\begin{equation*}
	\Upsilon_1 = \lbrace\, u_x - v_y = 0, \,\, u_y + v_x = 0, \,\,
	u_t=v_t=\dens_x=\dens_y=\entr_x=\entr_y  = 0 \,\rbrace,
	\end{equation*}
	\[
	\Upsilon_2 = \lbrace\,J_3J_5^2(J_3J_4-J_5^2)=0 \,\rbrace.
	\]
	The set $\Upsilon_1$ contains singular points that have five-dimensional orbits.
	The set $\Upsilon_2$ contains points where differential invariants $J_1,J_4,\ldots,J_{10}$ are dependent.
\end{proposition}

It is easy to check that codimension of regular $\LieAlgebra{g_{m}}$-orbits is equal to 12. 
The proofs of the following theorems can be found in \cite{Duyunova2017-4}.
\begin{theorem} \cite{Duyunova2017-4}
	The field of the first order kinematic invariants 
	is generated by invariants $\dens,\entr,J_1,\ldots,J_{10}$. 
	These invariants separate regular $\LieAlgebra{g_m}$-orbits.
\end{theorem}
\begin{theorem} \cite{Duyunova2017-4}
	The following derivations 
	\[
	\nabla_1 = \totalDiff{t}+u\totalDiff{x}+v\totalDiff{y},\quad
	\nabla_2 = \dens_x\totalDiff{x}+\dens_y\totalDiff{y},\quad
	\nabla_3 = \entr_x\totalDiff{x}+\entr_y\totalDiff{y}
	\]
	are $\LieAlgebra{g_{m}}$-invariant. They are linear independent if 
	\[
	\dens_x\entr_y-\dens_y\entr_x\neq 0.
	\]
\end{theorem}

The bundle $\pi_{2,1}:\systemEk{2}\rightarrow \systemEk{1}$ has rank 18 and by applying derivations $\nabla_1, \nabla_2, \nabla_3$ to the kinematic invariants $J_1,J_2,\ldots,J_{10}$ we get 30 kinematic invariants. Straightforward computations show that among these invariants 18 are always independent (see  \textit{http://d-omega.org}).

Therefore, beginning with order $k=1$ dimensions of regular orbits are equal to $\dim \LieAlgebra{g_m}=6$.

Moreover, the number of independent invariants of pure  order $k$ (the Hilbert function) is equal to 
$H(k) = 7k +4$ for $k\geq 2$, and $H(0)=2$,  $H(1)=10$.

The corresponding Poincar\'{e} function ie equal to
\begin{equation*}
P(z) = \frac{2+6z-z^3}{(1-z)^2}.
\end{equation*}
\begin{theorem} \cite{Duyunova2017-4}
	The field of kinematic invariants is generated by the invariants $\dens,\entr$ of order zero, invariants $J_1,J_2,\ldots,J_{10}$ of order one and by the invariant derivations  $\nabla_1, \nabla_2, \nabla_3$.  This field separates regular orbits.
\end{theorem}

\subsubsection*{The field of Navier--Stokes invariants}

Here we consider the case when the thermodynamic state admits a one-dimensional symmetry algebra generated by the vector field
\begin{equation*}
A =  \xi_1 X_7 + \xi_2 X_8 + \xi_3 X_9 + \xi_4 X_{10} .
\end{equation*}

Note that, the {$\LieAlgebra{g_{m}}$ invariant derivations $\nabla_1$, $\nabla_2$, $\nabla_3$ do not commute with the thermodynamic symmetry $A$. 
Moreover, the action of the thermodynamic vector field $A$ on the field of kinematic invariants is given by the following derivation
\[
(\xi_4-2\xi_3)\dens\dir{\dens} + \xi_1 \dir{\entr} - \xi_4J_1\dir{J_1} - \xi_4J_2\dir{J_2} - 2J_3(3\xi_3-\xi_4)\dir{J_3}  -
\]
\[
- 2\xi_3J_4\dir{J_4}  + J_5(\xi_4-4\xi_3)\dir{J_5} -\xi_4 J_6\dir{J_6} -
\]
\[
-4\xi_3J_7\dir{J_7} + J_{8}(\xi_4-6\xi_3)\dir{J_{8}} -2\xi_4J_{9}\dir{J_{9}} - (\xi_4+2\xi_3) J_{10}\dir{J_{10}}    .
\]
	
Finding the first integrals of this vector field we get the basic Navier-Stokes  invariants of the first order. The following result is valid for general $\xi$'s and the special cases can be found in \cite{Duyunova2017-4}.
\begin{theorem} \cite{Duyunova2017-4}
The field of the Navier--Stokes  differential invariants for the thermodynamic states admitting a one-dimensional symmetry algebra is generated by the differential invariants 
\[
\frac{\xi_1}{2\xi_3-\xi_4}\ln\dens+\entr  , \quad 
J_1 \dens^{\frac{\xi_4}{\xi_4-2\xi_3}}, 
\]
\[
\frac{J_2}{J_1}, \quad
\frac{J_3}{\dens^3J_1},  \quad
\frac{J_4}{\dens J_1}, \quad 
\frac{J_5}{\dens^2 J_1}, \quad
\frac{J_6}{J_1}, \quad
\frac{J_{7}}{\dens^3J_1^2}, \quad
\frac{J_8}{\dens^2J_1^2}, \quad 
\frac{J_{9}}{J_1^2}, \quad 
\frac{J_{10}}{\dens J_1^2}
\]
 of the first order and by the invariant derivations 
\[
\dens^{\frac{\xi_4}{\xi_4-2\xi_3}} \nabla_1,\quad
\dens^{\frac{2\xi_3}{\xi_4-2\xi_3}-1} \nabla_2,\quad
\dens^{\frac{2\xi_3}{\xi_4-2\xi_3}} \nabla_3 .
\] 
This field separates regular orbits.
\end{theorem}
	
Consider the case when the thermodynamic state admits a commutative two-dimensional symmetry algebra generated by the vector fields 
	$A=\sum\limits_{i=1}^6 \mu_iX_{i+6}$, $B=\sum\limits_{i=1}^6 \eta_iX_{i+6}$, then $\mu$'s and $\eta$'s satisfy relation 	
\begin{equation*}
	\eta_2\mu_4-\eta_4\mu_2=0.
\end{equation*}
	
Using similar computations we get.	
\begin{theorem} \cite{Duyunova2017-4}
The field of Navier--Stokes differential invariants for the thermodynamic states admitting a com\-mu\-ta\-tive two-dimensional symmetry algebra is generated by the differential invariants 
		\[
		J_1 \dens^{\varsigma_1 } e^{\varsigma_2 \entr}   , \quad
		\frac{J_2}{J_1}, \quad 
		\frac{J_3}{\dens^3J_1}, \quad
		\frac{J_4}{\dens J_1}, \quad 
		\frac{J_5}{\dens^2 J_1}, \quad
		\frac{J_6}{J_1}, \quad 
		\frac{J_7}{\dens^3J_1^2}, \quad 
		\frac{J_{8}}{\dens^2J_1^2}, \quad 
		\frac{J_{9}}{J_1^2}, \quad 
		\frac{J_{10}}{\dens J_1^2}
		\]
		of the first order and by the invariant derivations 
		\[
		\dens^{\varsigma_1 } e^{\varsigma_2 \entr}  \, \nabla_1,\quad
		\dens^{\varsigma_1-2 } e^{\varsigma_2 \entr}  \, \nabla_2,\quad
		\dens^{\varsigma_1-1 } e^{\varsigma_2 \entr}  \, \nabla_3,
		\]
		where
		\[
		\varsigma_1 = \frac{ \eta_4\mu_1-\eta_1\mu_4  }{2(\eta_1\mu_3-\eta_3\mu_1 ) +\eta_4\mu_1-\eta_1\mu_4   }, \quad
		\varsigma_2= \frac{ (\eta_4\mu_3-\eta_3\mu_4 ) }{2(\eta_1\mu_3-\eta_3\mu_1 ) +\eta_4\mu_1-\eta_1\mu_4   }.
		\]
		This field separates regular orbits.
\end{theorem}


\subsection{3D-flows}

Consider the Navier--Stokes system \eqref{eq:NS} in a space $M=\mathbb{R}^3$ equipped  with the coordinates $(x,y,z)$ and the standard metric $g=dx^2+dy^2+dz^2$. 

The velocity field of the flow has the form 
$\bvec{u}=u(t,x,y,z)\,\dir{x} +v(t,x,y,z)\,\dir{y}+w(t,x,y,z)\,\dir{z}$, 
the pressure $\press$, the density $\dens$, the temperature $\temp$ and the entropy $\entr$ are the functions of time and space with the coordinates $(t,x,y,z)$.

The vector of gravitational acceleration is of the form $\bvec{g}=g\,\dir{z}$.


\subsubsection{Symmetry Lie algebra}

First of all we consider the Lie algebra $\LieAlgebra{g}$ generated by the following vector fields on the manifold  $\JetSpace{0}{\pi}$ 
\begin{equation*} 
\begin{aligned} 
&X_1=\dir{x},\qquad
X_4=-y\,\dir{x}+x\,\dir{y}-v\,\dir{u}+u\,\dir{v}, \phantom{hjkhkjhkhgoo.}
\quad    \\ 
&X_2=\dir{y},\qquad 
X_5=\left(\frac{gt^2}{2}-z\right)\dir{x}+x\,\dir{z}+\left(gt-w\right)\dir{u}+u\,\dir{w}, \quad \\
&X_3=\dir{z},\qquad 
X_6=\left(\frac{gt^2}{2}-z\right)\dir{y}+y\,\dir{z}+\left(gt-w\right)\dir{v}+v\,\dir{w},  \quad \\
&X_7=t\,\dir{x}+\dir{u},\qquad X_{10}=\dir{t},\\
&X_8=t\,\dir{y}+\dir{v}, \qquad X_{11}= \dir{\entr}, \\
&X_9=t\,\dir{z}+\dir{w}, \qquad X_{12} = \dir{\press}, \\
&X_{13}=x\,\dir{x}+y\,\dir{y}-\left(\frac{gt^2}{2}-z\right)\dir{z}+u\,\dir{u}+v\,\dir{v}-\left(gt-w\right)\dir{w}
-2\dens\,\dir{\dens}+2\temp\,\dir{\temp},\\
&X_{14}=t\,\dir{t} +gt^2\dir{z}-u\,\dir{u}-v\,\dir{v}+\left(2gt-w\right)\dir{w}
+\dens\,\dir{\dens}-\press\,\dir{\press}-2\temp\,\dir{\temp}
\end{aligned} 
\end{equation*}
and the Lie algebra $\LieAlgebra{h}$ generated by the vector fields
\begin{equation*}
Y_1=\dir{\entr} ,  \quad    Y_2=\dir{\press}, \quad 
Y_3=\dens\,\dir{\dens}-\temp\,\dir{\temp},  \quad    
Y_4=\press\,\dir{\press}+\temp\,\dir{\temp}.
\end{equation*}

The pure geometric part is represented by the algebra $\LieAlgebra{g_{m}}=\left\langle X_1,X_2, \ldots, X_{10} \right\rangle $ with respect to the group of motions, Galilean transformations and time shifts. 

In order to describe the pure thermodynamic part, we consider the Lie subalgebra $\LieAlgebra{h_{t}}$ of the algebra $\LieAlgebra{h}$ that preserves the thermodynamic state \eqref{eq:Therm}.
\begin{theorem} \cite{Duyunova2017-5} 
	A Lie algebra  $\LieAlgebra{g_{sym}}$ of symmetries of the Navier--Stokes system of differential equations in 3-dimensional space coincides with 
	\[
	\vartheta^{-1}(\LieAlgebra{h_{t}}).
	\]
\end{theorem}

\subsubsection{Symmetry classification of states}

The Lie algebra generated by the vector fields $Y_1, \ldots, Y_4$ coincides with the Lie algebra of the thermodynamic symmetries of the Navier--Stokes system on a plane. 

Thus the classification of the thermodynamic states or Lagrangian surfaces  $\lagrangianSurface$ depending on the dimension of the symmetry algebra $\LieAlgebra{h_{t}}\subset \LieAlgebra{h}$ is the same as the classification presented in the previous section (2D-flows).


\subsubsection{Differential invariants}

\subsubsection*{The field of kinematic invariants}

First of all, we observe that the functions $
\dens$ and $\entr$
(as well as $\press$ and $\temp$) generate all $\LieAlgebra{g_{m}}$-invariants of order zero. 

Let fix the $0$ point with coordinates $(0,\ldots,0)\in \JetSpace{0}{\pi}$ and consider the isotropy group of this point. It is easy to check that this group is isomorphic to the rotation group $SO(3)$. 

Then consider the following elements
\[
\veltg=
\begin{pmatrix}
u_t \\
v_t \\
w_t-g
\end{pmatrix}, \quad
\nabla \dens=
\begin{pmatrix}
\dens_x \\
\dens_y \\
\dens_z
\end{pmatrix},  \quad
\nabla\entr=
\begin{pmatrix}
\entr_x \\
\entr_y \\
\entr_z
\end{pmatrix},
 \quad
\velxyz=
\begin{pmatrix}
u_x & u_y & u_z \\
v_x & v_y & v_z \\
w_x & w_y & w_z
\end{pmatrix}
\] 
and suppose that first three vectors are linearly independent.

Note that, the group $SO(3)$ acts on the matrix $\velxyz$ by conjugacy:  $\velxyz \rightarrow R\velxyz R^{-1}$, where $R\in SO(3)$.

Moreover, the action of the rotation group $SO(3)$  preserves the dot products of the vectors $\veltg$, $\nabla \dens$ and $\nabla\entr$.

Let $\mathrm{H} = (\veltg,\nabla{\dens},\nabla{\entr} )$ be a matrix with $\det \mathrm{H}\neq 0$, then the elements of the product $\mathrm{H}^{-1}\velxyz \mathrm{H}$ are 9 functions, which are invariant under the action of the rotation group.

Therefore, we have 15 independent invariants of the first order at the point  $(0,\ldots,0)$.

Denote by $\tau$ the following transformation:
\[
\begin{aligned}
&t \rightarrow t-t_0  , \qquad
&x \rightarrow x-x_0-u_0(t-t_0), \qquad &u \rightarrow u-u_0,\\
& \dens \rightarrow \dens, \qquad
&y \rightarrow y-y_0-\,v_0(t-t_0), \qquad  &v \rightarrow v-v_0,\\
& \entr \rightarrow \entr, \qquad
&z \rightarrow z-z_0-w_0(t-t_0), \qquad &w \rightarrow w-w_0.
\end{aligned}
\] 

Obviously, $\tau$ is a symmetry of the equation
$\systemEk{}$, which maps the point
$(t_0,x_0,y_0,z_0$ $u_0,v_0,w_0,$ $\dens_0,\entr_0)$ to the point $0$.

Applying the prolongation of $\tau$ to the invariants (to the dot products and the elements of the matrix $\mathrm{H}^{-1}\velxyz \mathrm{H}$) we get 15 kinematic invariants of the first order. 

The proofs of the following two theorems can be found in \cite{Duyunova2017-5}.
\begin{theorem} \cite{Duyunova2017-5}
	The field of the first order kinematic invariants 
	is generated by the invariants $\dens, \entr$ and by the invariants 
	\begin{align}\label{eq:invars3d}
	\begin{split}
	&\entr_t + \entr_x u + \entr_y v + \entr_z w, \quad  \\
	&(\nabla{\dens})^2, \quad (\nabla{\entr})^2, \quad \nabla{\dens}\cdot\nabla{\entr}, 
	\quad (\veltg)^2, \quad \nabla{\dens}\cdot\veltg, \quad \nabla{\entr}\cdot\veltg,\\
	&(\mathrm{H}^{-1}\velxyz \mathrm{H})_{ij},
	\end{split}
	\end{align}
	transformed by $\tau$, if  $\det \mathrm{H} \neq 0 $. These invariants separate regular $\LieAlgebra{g_m}$-orbits.
\end{theorem}
\begin{theorem} \cite{Duyunova2017-5}
	The following derivations 
	\[
	\begin{aligned}
	&\phantom{kjhkkjhkjdf}
	\nabla_1 = \totalDiff{t}+u\totalDiff{x}+v\totalDiff{y}+w\totalDiff{z},\\
	&\nabla_2 = \dens_x\totalDiff{x}+\dens_y\totalDiff{y}+\dens_z\totalDiff{z},\qquad
	\nabla_3 = \entr_x\totalDiff{x}+\entr_y\totalDiff{y}+\entr_z\totalDiff{z} , \\
	& \nabla_4 = (\dens_y\entr_z-\dens_z\entr_y)\totalDiff{x}-(\dens_x\entr_z-\dens_z\entr_x)\totalDiff{y}+(\dens_x\entr_y-\dens_y\entr_x)\totalDiff{z} 
	\end{aligned} 
	\]
	are $\LieAlgebra{g_{m}}$-invariant. They are linearly independent if 
	\[
	\left| \begin{array}{ccc}
	\dens_x & \dens_y & \dens_z \\
	\entr_x & \entr_y & \entr_z \\
	\dens_y\entr_z-\dens_z\entr_y & \dens_x\entr_z-\dens_z\entr_x & \dens_x\entr_y-\dens_y\entr_x
	\end{array} \right|
	\neq 0.
	\]
\end{theorem}

The bundle $\pi_{2,1}:\systemEk{2}\rightarrow \systemEk{1}$ has rank 42 and by applying the derivations $\nabla_i$, $i=1,\ldots,4$ to the kinematic invariants \eqref{eq:invars3d} we get 64 kinematic invariants. Straightforward computations show that among these invariants 42 are always independent (see  \textit{http://d-omega.org}).

Therefore, starting with the order $k=1$ dimensions of regular orbits are equal to $\dim \LieAlgebra{g_m}=10$.

The Hilbert function of the  $\LieAlgebra{g_m}$-invariants field (the number of independent invariants of pure order $k$) is equal to
\[
H(k) = \frac{9}{2}k^2 + \frac{19}{2} k + 5
\]
for $k\geq 2$, and $H(0)=2$,  $H(1)=16$.

The corresponding Poincar\'{e} function has the form
\begin{equation*}
P(z) = \frac{2+10z-6z^3+3z^4}{(1-z)^3}.
\end{equation*}

Summarizing, we get the following result.
\begin{theorem} \cite{Duyunova2017-5}
	The field of kinematic invariants is generated by the invariants $\dens, \entr$ of order zero, by the invariants \eqref{eq:invars3d} of order one (with transformation $\tau$) and by the invariant derivations  $\nabla_i$, $i=1,\ldots,4$. This field separates the regular orbits.
\end{theorem}

\subsubsection*{The field of Navier--Stokes invariants }

Consider the case when the equations of thermodynamic state  admits a one-dimensional symmetry algebra generated by the vector field
\begin{equation*}
A =  \xi_1 X_{11} + \xi_2 X_{12} + \xi_3 X_{13} + \xi_4 X_{14} .
\end{equation*}

For general $\xi$'s we have the following result. The particular cases are considered in \cite{Duyunova2017-5}.
\begin{theorem} \cite{Duyunova2017-5}
The field of the Navier--Stokes  differential invariants for the thermodynamic states admitting a one-dimensional symmetry algebra is generated by the differential invariants  
\[
\begin{aligned}
&\frac{\xi_1}{2\xi_3-\xi_4}\ln\dens+\entr  , \qquad 
\dens^{\frac{\xi_4}{\xi_4-2\xi_3}} \nabla_1\,\entr, \qquad
\frac{\left(\nabla\dens \right)^2}{\dens^3\, \nabla_1\entr}, \qquad
\frac{\left(\nabla\entr \right)^2}{\dens\, \nabla_1\entr}, \qquad
\frac{\nabla\dens\cdot \nabla\entr}{\dens^2\, \nabla_1\entr},
\\ 
&
\frac{\dens \,(\veltg)^2}{ \left(\nabla_1\entr\right)^3}, \qquad
\frac{\nabla\dens\cdot \veltg}{\dens\, (\nabla_1\entr)^2}, \qquad
\frac{\nabla\entr\cdot \veltg}{ (\nabla_1\entr)^2}, \qquad
\frac{J_{11}}{\nabla_1\entr}, \qquad
\frac{J_{12}}{\dens^2}, \qquad 
\frac{J_{13}}{\dens}, \qquad \\
&\frac{\dens^2 J_{21}}{(\nabla_1\entr)^2}, \qquad
\frac{J_{22}}{\nabla_1\entr}, \qquad
\frac{\dens J_{23}}{\nabla_1\entr}, \qquad
\frac{\dens J_{31}}{(\nabla_1\entr)^2}, \qquad
\frac{\dens J_{32}}{\dens\,\nabla_1\entr}, \qquad
\frac{\dens J_{3}}{\nabla_1\entr}
\end{aligned}
\]
of the first order and by the invariant derivations  
\[
	\dens^{\frac{\xi_4}{\xi_4-2\xi_3}} \nabla_1,\quad
	\dens^{-\frac{\xi_4-4\xi_3}{\xi_4-2\xi_3}} \nabla_2,\quad
	\dens^{\frac{2\xi_3}{\xi_4-2\xi_3}} \nabla_3,\quad 
	\dens^{-\frac{\xi_4-5\xi_3}{\xi_4-2\xi_3}} \nabla_4,
\]
	here we denote by $J_{ij}$ the elements of the matrix $\mathrm{H}^{-1}\velxyz \mathrm{H}$.  This field separates the regular orbits.
\end{theorem}
	
Consider the case when the thermodynamic state admits a commutative two-dimensional symmetry algebra generated by the vector fields 
$A=\sum\limits_{i=1}^6 \mu_iX_{i+10}$, $B=\sum\limits_{i=1}^6 \eta_iX_{i+10}$. 
\begin{theorem} \cite{Duyunova2017-5}
The field of the Navier--Stokes differential invariants for the thermodynamic states admitting a com\-mu\-ta\-tive two-dimensional symmetry algebra is generated by the differential invariants 
\begin{equation*}
		\begin{aligned}
		&\dens^{\varsigma_1 } e^{\varsigma_2 \entr} \nabla_1\entr    , \qquad
		\frac{(\nabla\dens)^2}{\dens^3\,\nabla_1\entr}, \qquad
		\frac{(\nabla\entr)^2}{\dens\,\nabla_1\entr},\qquad
		\frac{\nabla\dens\cdot\nabla\entr}{\dens^2\,\nabla_1\entr}, \\
		&\frac{\dens (\veltg)^2}{(\nabla_1\entr)^3}, \qquad
		\frac{\nabla\dens\cdot\veltg}{\dens(\nabla_1\entr)^2}, \qquad
		\frac{\nabla\entr\cdot\veltg}{(\nabla_1\entr)^2}, \qquad
		\frac{J_{11}}{\nabla_1\entr}, \qquad 
		\frac{J_{12}}{\dens^2}, \qquad  
		\frac{J_{13}}{\dens},   \\
		&\frac{\dens^2 J_{21}}{(\nabla_1\entr)^2}, \qquad
		\frac{J_{22}}{\nabla_1\entr}, \qquad
		\frac{\dens\, J_{23}}{\nabla_1\entr}, \qquad
		\frac{\dens\, J_{31}}{(\nabla_1\entr)^2}, \qquad
		\frac{J_{32}}{\dens\,\nabla_1\entr}, \qquad
		\frac{J_{33}}{\nabla_1\entr}
		\end{aligned}
\end{equation*}
of the first order and by the invariant derivations 
		\[
		\dens^{\varsigma_1 } e^{\varsigma_2 \entr} \, \nabla_1,\qquad
		\dens^{\varsigma_1 -2 } e^{\varsigma_2 \entr} \, \nabla_2,\qquad
		\dens^{\varsigma_1 -1 } e^{\varsigma_2 \entr}  \, \nabla_3, \qquad
		\dens^{\frac{3}{2} \varsigma_1 -\frac{5}{2} } e^{\frac{3}{2}\varsigma_2 \entr} \, \nabla_4
		\]
where $J_{ij}$ are the elements of the matrix $\mathrm{H}^{-1}\velxyz \mathrm{H}$ and 
		\[
		\varsigma_1 = \frac{ \eta_4\mu_1-\eta_1\mu_4  }{2(\eta_1\mu_3-\eta_3\mu_1 ) +\eta_4\mu_1-\eta_1\mu_4   }, \quad
		\varsigma_2= \frac{ 2(\eta_4\mu_3-\eta_3\mu_4 ) }{2(\eta_1\mu_3-\eta_3\mu_1 ) +\eta_4\mu_1-\eta_1\mu_4   }.
		\]
		This field separates the regular orbits.
\end{theorem}

\subsection{Flows on a sphere}
Consider Navier--Stokes system \eqref{eq:E} on a two-dimensional unit sphere $M=S^2$  with the metric $g=\sin^2y\,dx^2+dy^2$ in the spherical coordinates.

The velocity field of the flow has the form 
$\bvec{u}=u(t,x,y)\,\dir{x} +v(t,x,y)\,\dir{y}$, 
the pressure $\press$, the density $\dens$, the temperature $\temp$ and the entropy $\entr$ are the functions of time and space with the coordinates $(t,x,y)$.

Here we consider the flow without any external force field, so $\bvec{g}={0}$.

\subsubsection{Symmetry Lie algebra}

To describe the Lie algebra of symmetries of the Navier--Stokes system we consider the Lie algebra $\LieAlgebra{g}$ generated
by the following vector fields on the manifold $\JetSpace{0}{\pi}$: 
\begin{equation*}
\begin{aligned} 
&X_1 = \dir{ t}, \qquad
X_2 = \dir{ x}, \qquad  \\
&X_{3} = \frac{\cos x}{\tan y} \,\dir{ x} + \sin x \,\dir{ y} -
\left( \frac{\sin x}{\tan y}\,u + \frac{\cos x}{\sin^2 y} \,v \right) \dir{ u} + u\cos x \,\dir{ v}  ,  \\
&X_{4} =   \frac{\sin x}{\tan y} \,\dir{ x} - \cos x \,\dir{ y} +
\left( \frac{\cos x}{\tan y}\,u - \frac{\sin x}{\sin^2 y} \,v \right) \dir{ u} + u\sin x \,\dir{ v}  ,  \\
&X_{5} =  \dir{ \entr} , \qquad  X_{6} =  \dir{ \press} , \\
&X_{7} = t\,\dir{ t} - u\,\dir{ u} - v\,\dir{ v} - \press\,\dir{ \press} + \dens\,\dir{ \dens} -2 \temp\,\dir{ \temp}  ,
\end{aligned} 
\end{equation*}
and denote by $\LieAlgebra{h}$ the Lie algebra generated by the vector fields
\begin{equation*}
Y_1 = \dir{ \entr}, \qquad
Y_2 = \dir{ \press}, \qquad
Y_3 =  \press\,\dir{ \press} - \dens\,\dir{ \dens} +2 \temp\,\dir{ \temp}.
\end{equation*}


Transformations corresponding to elements of the algebra $\LieAlgebra{g_{m}}=\left\langle X_1,X_2, X_3, X_4\right\rangle $ (the pure geometric part) are generated by sphere motions and time shifts: $\LieAlgebra{g_{m}} = \LieAlgebra{so}(3,\mathbb{R})\oplus \mathbb{R}$.

For describing the pure thermodynamic part we consider  the Lie subalgebra $\LieAlgebra{h_{t}}$ of algebra $\LieAlgebra{h}$ that preserves the thermodynamic state \eqref{eq:Therm}.
\begin{theorem} \cite{Duyunova2017-6} 
	The Lie algebra $\LieAlgebra{g_{sym}}$ of point symmetries of the Navier--Stokes system of differential equations on a sphere coincides with 
	\[
	\vartheta^{-1}(\LieAlgebra{h_{t}}).
	\]
\end{theorem}


\subsubsection{Symmetry classification of states}

Here we consider the thermodynamic states or Lagrangian surfaces  $\lagrangianSurface$ (compare with the plane case)
with a one-dimensional symmetry algebra $\LieAlgebra{h_{t}}\subset \LieAlgebra{h}$ .

Cases when the thermodynamic states admit two or three-dimensional symmetry algebra are not interesting or have no physical meaning.
	
Let $\dim \LieAlgebra{h_{t}}=1 $ and let $Z = \sum\limits_{i=1}^3 \lambda_i Y_i$ be a basis vector in this algebra.

Then the thermodynamic state or the surface $\lagrangianSurface$ is the solution of the PDE system
\begin{equation*}
\left\{
\begin{aligned}
&\lambda_3\dens\,{\energy}_{\dens\dens} - \lambda_1{\energy}_{\dens \entr}  + 3\lambda_3 {\energy}_{\dens}  + \frac{\lambda_2}{\dens^2} =0, \\
&\lambda_1{\energy}_{\entr \entr} - \lambda_3\dens\,{\energy}_{\dens \entr} -2\lambda_3\,{\energy}_{ \entr} =0,
\end{aligned} \right. 
\end{equation*}
that is formally integrable and compatible.

Solving this system for general parameters $\lambda$ (some special cases can be found in \cite{Duyunova2017-6}) we find expressions for the pressure and  the temperature. 
Adding the admissibility conditions for this case we get the following result.
\begin{theorem}
	The thermodynamic states admitting a one-dimensional symmetry algebra  have the form 
	\[
\press = \frac{1}{\dens} \left( \frac{\lambda_1}{\lambda_3}F^{\prime} -2F\right)    - \frac{\lambda_2 }{\lambda_3}  ,\quad
\temp =  \frac{F^{\prime}}{\dens^2}  , \quad
F=F\left( \entr + \frac{\lambda_1}{\lambda_3}\ln \dens \right),
	\]
	where $F$ is an arbitrary function and
	\[
F^{\prime}>0, \quad
\left( \frac{\lambda_1}{\lambda_3}\right)^2  F^{\prime\prime} -3\frac{\lambda_1}{\lambda_3}F^{\prime} +2F >0,
 \quad    F^{\prime\prime}\left( \frac{\lambda_1 }{\lambda_3}F^{\prime} +2F\right) - 4(F^{\prime})^2 
>0.
	\]
\end{theorem}

\subsubsection{Differential invariants}

\subsubsection*{The field of kinematic invariants}

First of all, the pressure
$\dens$, the entropy $\entr$ and $g(\bvec{u},\bvec{u})$ 
(as well as $\press$ and $\temp$) generate all $\LieAlgebra{g_{m}}$-invariants of order zero.

Consider two vector fields $\bvec{u}$ and $\tilde{\bvec{u}}$ such that $g(\bvec{u},\tilde{\bvec{u}})=0$ and $g(\bvec{u},\bvec{u})=g(\tilde{\bvec{u}},\tilde{\bvec{u}})$. Writing the acceleration vector  with respect to the vectors $\bvec{u}$ and $\tilde{\bvec{u}}$ we obtain two invariants of the first order.  Further writing the operator  $d_{\nabla}\bvec{u}$ with respect to these vectors
as the sum of its symmetric and antisymmetric parts we obtain another four invariants of the first order. Thus we get six invariants:
\begin{equation}\label{eq:invarsNSs}
\begin{aligned}
&J_1 = (uv_t-vu_t)\sin y, \qquad J_2 = uu_t\sin^2y + vv_t, \\
&J_3 = u_x + v_y + v\cot y, \qquad 
J_4 = u_y\sin y - \frac{v_x}{\sin y} +2 u \cos y, \\
&J_5 = (u(u_xv-v_xu)+v(u_yv-v_yu))\sin y +u \cos y (u^2\sin^2y+2v^2), \\
&J_6 = v(u_xv-v_xu)-u(u_yv-v_yu)\sin^2 y  + v^3\cot y.
\end{aligned}
\end{equation}

The proofs of the following theorems can be found in \cite{Duyunova2017-6}.
\begin{theorem} \cite{Duyunova2017-6}
	The following derivations 
	\[
	\nabla_1 = \totalDiff{t} ,\quad
	\nabla_2 = \frac{\dens_x}{\sin^2 y}
	\totalDiff{x}+\dens_y\totalDiff{y},\quad
	\nabla_3 = \frac{\entr_x}{\sin^2 y} \totalDiff{x}+\entr_y\totalDiff{y}
	\]
	are $\LieAlgebra{g_{m}}$-invariant. They are linearly independent if 
	\[
	\dens_x\entr_y-\dens_y\entr_x\neq 0.
	\]
\end{theorem}
\begin{theorem} \cite{Duyunova2017-6}
	The field of the first order kinematic invariants 
	is generated by the invariants $\dens,\, \entr,\, g(\bvec{u},\bvec{u})$, \eqref{eq:invarsNSs} and 
	\begin{equation}\label{eq:invarsNSs2}
	\nabla_1\dens, \quad \nabla_1\entr, \quad
	\nabla_2\dens, \quad  \nabla_2\entr, \quad \nabla_3\entr.
	\end{equation}
	These invariants separate regular $\LieAlgebra{g_m}$-orbits.
\end{theorem}

The bundle $\pi_{2,1}:\systemEk{2}\rightarrow \systemEk{1}$ has rank 18,
and by applying derivations $\nabla_1, \nabla_2, \nabla_3$ to the kinematic invariants \eqref{eq:invarsNSs} and \eqref{eq:invarsNSs2} we get 33 kinematic invariants.
Straightforward computations show that among these invariants 18 are independent (see  \textit{http://d-omega.org}).

Therefore, starting with the order $k=1$ dimensions of regular orbits are equal to $\dim \LieAlgebra{g_m}=4$.

Moreover, the number of independent
invariants of pure order $k$ (the Hilbert function) is equal $H(k)=7k+4$ for $k\geq 1$, and $H(0)=3$.

The corresponding Poincar\'{e} function is 
\begin{equation*}
P(z) = \frac{3+5z-z^2}{(1-z)^2}.
\end{equation*}

\begin{theorem} \cite{Duyunova2017-6}
	The field of the kinematic invariants is generated by the invariants $\dens, \,  \entr, \, g(\bvec{u},\bvec{u})$ of order zero,
	by the invariants \eqref{eq:invarsNSs} and \eqref{eq:invarsNSs2} of order one and by the invariant derivations  $\nabla_1, \nabla_2, \nabla_3$.
	This field separates regular orbits.
\end{theorem}

\subsubsection*{The field of Navier--Stokes invariants }

Now, we find differential invariants of the Navier--Stokes system in the case when the thermodynamic state  $\lagrangianSurface$  admits a one-dimensional symmetry algebra generated by the vector field
\begin{equation*}
A =  \xi_1 X_5 + \xi_2 X_6 + \xi_3 X_7  .
\end{equation*}
\begin{theorem} \cite{Duyunova2017-6}
	The field of the Navier--Stokes differential invariants for thermodynamic states
	admitting a one-dimensional symmetry algebra is generated by
	the differential invariants 
	\begin{equation*}
	\begin{aligned}
	& \entr - \frac{\xi_1}{\xi_3}\ln \dens   , \quad 
	\dens^2g(\bvec{u},\bvec{u}), \quad
	\dens^3J_1, \quad 
	\dens^3J_2, \quad
	\dens J_3,  \quad
	\dens J_4,\\
	&\dens^3 J_5, \quad 
	\dens^3 J_6, \quad
	\nabla_1\dens , \quad
	\dens\nabla_1\entr, \qquad
	\frac{\nabla_2\dens}{\dens^2}, \quad 
	\frac{\nabla_2\entr}{\dens}, \quad
	\nabla_3\entr
	\end{aligned}
	\end{equation*}
	of the first order and by the invariant derivations 
	\begin{equation*} 
	\dens \nabla_1,\quad
	\dens^{-1} \nabla_2,\quad
	\nabla_3 .
	\end{equation*}
	This field separates regular orbits.
\end{theorem}

This theorem is valid for general $\xi$'s. For special cases see \cite{Duyunova2017-6}.


\subsection{Flows on a spherical layer}
Consider the Navier--Stokes system \eqref{eq:NS} on a spherical layer $M=S^2\times \mathbb{R}$ with the coordinates $(x,y,z)$ and the metric 
\[
g=\frac{4}{(x^2+y^2+1)^2}(dx^2+dy^2) +dz^2.
\]

The velocity field of the flow has the form 
$\bvec{u}=u(t,x,y,z)\,\dir{x} +v(t,x,y,z)\,\dir{y}+w(t,x,y,z)\,\dir{z}$, 
the pressure $\press$, the density $\dens$, the temperature $\temp$ and the entropy $\entr$ are the functions of time and space with the coordinates $(t,x,y,z)$.

The vector of gravitational acceleration is of the form $\bvec{g}=g\,\dir{z}$.

\subsubsection{Symmetry Lie algebra}

Consider the Lie algebra $\LieAlgebra{g}$ generated
by the following vector fields on the manifold $\JetSpace{0}{\pi}$: 
\begin{equation*}
\begin{aligned} 
&X_1 = \dir{ t}, \qquad
X_2 = \dir{ z}, \qquad 
X_3 = t\,\dir{ z}+\dir{ w}, \qquad\\
&X_4 = y\,\dir{ x}-x\,\dir{y}+v\,\dir{u}-u\,\dir{v}, \\
&X_{5} = xy \,\dir{ x} -\frac{1}{2} (x^2-y^2-1)\dir{ y} +
\left( xv + yu \right) \dir{ u} - (xu-yv)\dir{ v}  ,  \\
&X_{6} =    \frac{1}{2}(x^2-y^2+1)\dir{ x} + xy \,\dir{ y} +
(xu-yv) \dir{ u} + (xv+yu)\dir{ v}  ,  \\
&X_{7} =  \dir{ \entr} , \qquad  X_{8} =  \dir{ \press} , \\
&X_{9} = t\,\dir{ t} +\mathrm{g}t^2\,\dir{ z} - u\,\dir{ u} - v\,\dir{ v} +(2\mathrm{g}t- w)\dir{ w} - \press\,\dir{ \press} + \dens\,\dir{ \dens} -2 \temp\,\dir{ \temp}  ,
\end{aligned} 
\end{equation*}
and denote by $\LieAlgebra{h}$ the Lie algebra generated by the vector fields
\begin{equation*}
Y_1 = \dir{ \entr}, \qquad
Y_2 = \dir{ \press}, \qquad
Y_3 =  \press\,\dir{ \press} - \dens\,\dir{ \dens} +2 \temp\,\dir{ \temp}.
\end{equation*}


Transformations corresponding to the elements of the algebra $\LieAlgebra{g_{m}}=\left\langle X_1,\ldots, X_6 \right\rangle $ (the pure geometric part) are compositions of sphere motions, Galilean transformations along the $z$ direction and time shifts. 

Let also $\LieAlgebra{h_{t}}$ be the Lie subalgebra of algebra $\LieAlgebra{h}$
that preserves the thermodynamic state \eqref{eq:Therm}.
\begin{theorem} \cite{Duyunova2017-7}
	The Lie algebra $\LieAlgebra{g_{sym}}$ of point symmetries of the Navier--Stokes system of differential equations on a spherical layer coincides with 
	\[
	\vartheta^{-1}(\LieAlgebra{h_{t}}).
	\]
\end{theorem}

\subsubsection{Symmetry classification of states}

The Lie algebra generated by the vector fields $Y_1, Y_2, Y_3$ coincides with the Lie algebra of thermodynamic symmetries of the Navier--Stokes system on a sphere. 
	
Thus the classification of thermodynamic states or Lagrangian surfaces  $\lagrangianSurface$ depending on the dimension of the symmetry algebra $\LieAlgebra{h_{t}}\subset \LieAlgebra{h}$ is the same as the classification presented in the previous section.


\subsubsection{Differential invariants}

\subsubsection*{The field of kinematic invariants}

First of all, the following functions
$
\dens, \, \entr,  \, g(\bvec{u},\bvec{u})-w^2
$
(as well as $\press$ and $\temp$) generate all $\LieAlgebra{g_{m}}$-invariants of order zero.

The proofs of the following theorems can be found in \cite{Duyunova2017-7}.
\begin{theorem} \cite{Duyunova2017-7}
	The following derivations 
	\[
	\nabla_1 = \totalDiff{z} ,\quad
	\nabla_2 = \totalDiff{t}+w\totalDiff{z},\quad
	\nabla_3 = u\totalDiff{x}+v\totalDiff{y},\quad
	\nabla_4 =  v \totalDiff{x}-u\totalDiff{y}  
	\]	
	are $\LieAlgebra{g_{m}}$-invariant. They are linearly independent if 
	\[
	u^2+v^2\neq 0.
	\]
\end{theorem}
\begin{theorem} \cite{Duyunova2017-7}
	The field of the first order kinematic invariants 
	is generated by the invariants $\dens,\, \entr,\, g(\bvec{u},\bvec{u})-w^2$ of order zero and by the invariants
	\begin{equation}\label{eq:invarsNSll}
	\begin{aligned}
	&\nabla_i\dens,  \quad
	\nabla_i\entr, \quad 
	\nabla_i(g(\bvec{u},\bvec{u})-w^2),\quad
	\nabla_i w,  \\
	&J_1 = u_z w_x+v_z w_y,   
	\quad J_2={(u_x-v_y)}^2+{(u_y+v_x)}^2,
	\quad J_3=\frac{u_t v_z-u_z v_t}{(x^2+y^2+1)^2}
	\end{aligned}
	\end{equation}
	of order one, where $i=1,\ldots,4$. 
	These invariants separate regular $\LieAlgebra{g_m}$-orbits.
\end{theorem}

The bundle $\pi_{2,1}:\systemEk{2}\rightarrow \systemEk{1}$ has rank 42,
and by applying derivations $\nabla_i$, $i=1,\ldots,4$ to the kinematic
invariants~\eqref{eq:invarsNSll} we get 88 kinematic invariants.
Straightforward computations show that among these invariants 42 are independent (see  \textit{http://d-omega.org}).

Therefore, starting with the order $k=1$ dimensions of regular orbits are equal to $\dim \LieAlgebra{g_m}=6$.

The  number of independent
invariants of pure order $k$ (the Hilbert function) is equal to
\[
H(k) = 5+\frac{19}{2}k +\frac{9}{2}k^2
\]
for $k\geq 1$, and $H(0)=3$.

The corresponding Poincar\'{e} function has the form
\begin{equation*}
P(z) = \frac{3+10z-6z^2+2z^3}{(1-z)^3}.
\end{equation*}
\begin{theorem} \cite{Duyunova2017-7}
	The field of the kinematic invariants is generated by
	the invariants $\dens, \,  \entr, \, g(\bvec{u},\bvec{u})-w^2$ of order zero,
	by the invariants~\eqref{eq:invarsNSll} of order
	one and by the invariant derivations  $\nabla_i$, $i=1,\ldots,4$.
	This field separates regular orbits.
\end{theorem}

\subsection{The field of Navier--Stokes invariants }
Consider the case when the thermodynamic state  $\lagrangianSurface$ 
admits a one-dimensional symmetry algebra generated by the vector field
\begin{equation*}
A =  \xi_1 X_7 + \xi_2 X_8 + \xi_3 X_9.
\end{equation*}

We do not consider cases of a two- or three-dimensional symmetry
algebra because they are not interesting from the
physical point of view.

For general $\xi$'s we have the following theorem. For the case $\xi_3=0$ we have basic invariants $\dens,\,g(\bvec{u},\bvec{u})-w^2$, \eqref{eq:invarsNSll} and invariant derivatives $\nabla_i,$ $i=1,\ldots,4$.
\begin{theorem} \cite{Duyunova2017-7}
	The field of the Navier--Stokes differential invariants for thermodynamic states
	admitting a one-dimensional symmetry algebra is generated by
	the differential invariants 
	\begin{equation*}
	\begin{aligned}
	& \entr - \frac{\xi_1}{\xi_3}\ln \dens   , \quad 
	\dens^2(g(\bvec{u},\bvec{u})-w^2), \quad
	\frac{\nabla_1\dens}{\dens}, \quad 
	\nabla_j\dens, \quad
	\nabla_1\entr,  \quad
	\dens\nabla_j\entr,\\
	&\dens^2 \nabla_1(g(\bvec{u},\bvec{u})-w^2), \quad 
	\dens^3 \nabla_j(g(\bvec{u},\bvec{u})-w^2), \quad\\
	&\dens\nabla_1w, \qquad
	\dens^2(\nabla_2 w - \mathrm{g}), \quad
	\dens^2\nabla_3w, \qquad
	\dens^2\nabla_4w, \qquad \\
	&\dens^2J_1, \quad 
	\dens^2J_2, \quad
	\dens^3J_3
	\end{aligned}
	\end{equation*}
	of the first order, here $j=2,3,4$, and by the invariant derivations 
	\begin{equation*}
	\nabla_1,\quad
	\dens \nabla_2,\quad
	\dens \nabla_3,\quad
	\dens \nabla_4.
	\end{equation*}
	This field separates regular orbits.
\end{theorem}

{\bf Acknowledgments.}
The research was partially supported by RFBR Grant No 18-29-10013.


\end{document}